\documentclass[a4paper,12pt]{article}
\usepackage{amsmath,amsfonts,graphicx,afterpage}
\renewcommand{\/}[1]{\mathbf{\boldsymbol{#1}}}
\usepackage{listings}
\usepackage{multirow}
\usepackage{pbox}
\usepackage{amsthm}
\usepackage{subfig}
\usepackage[sort]{natbib}
\bibpunct{(}{)}{;}{a}{,}{,} 


\usepackage{float}
\usepackage{color}
\usepackage{colortbl}
\newcommand{\var}{\text{var}}
\newcommand{\diag}{\text{diag}}
\newcommand{\ICS}{\text{ICS}}
\newcommand{\PP}{\text{PP}}
\newcommand{\kurt}{\text{kurt}}
\newcommand{\trunc}{\text{trunc}}
\usepackage[usenames,dvipsnames,svgnames,table]{xcolor}
\author{Fatimah Alashwali$^{(1)}$ and John Kent$^{(2)}$\\
$^{(1)}$ Department of Mathematical Science,\\ 
Princess Nourah bint Abdulrahman University\\Riyadh, Saudi Arabia\\
$^{(2)}$ Department of Statistics, University of Leeds, LS2 9JT
}
\title{The use of a common location measure in the invariant 
coordinate selection and projection pursuit}
\begin{document}
\maketitle
\begin{abstract}
  Invariant coordinate selection (ICS) and projection pursuit (PP) are
  two methods that can be used to detect clustering directions in
  multivariate data by optimizing criteria sensitive to
  non-normality. In particular, ICS finds clustering directions using
  a relative eigen-decomposition of two scatter matrices with
  different levels of robustness; PP is a one-dimensional variant of ICS.
  Each of the two scatter matrices includes an implicit or explicit
  choice of location.  However, when different measures of location
  are used, ICS and PP can behave counter-intuitively.  In this paper we
  explore this behavior in a variety of examples and propose a simple
  and natural solution: use the same measure of location for both
  scatter matrices.
\end{abstract}
\section{Introduction}
\label{sec:sec1}
Consider a multivariate dataset, given as an $n \times p$ data matrix
$X$, and suppose we want to explore the existence of any clusters.
One way to detect clusters is by projecting the data onto a lower
dimensional subspace for which the data are maximally non-normal.
Hence, methods that are sensitive to non-normality can be used to
detect clusters.

One set of methods based on this principle is invariant coordinate
selection (ICS), introduced by \citet{Tyler}, together with a
one-dimensional variant called projection pursuit (PP), introduced by
\citet{friedman1974}.  ICS involves the use of two scatter matrices,
$S_1 = S_1(X)$ and $S_2 = S_2(X)$ with $S_2$ chosen to be more robust than
$S_1$.
An eigen-decomposition of $S_2^{-1}S_1$ is carried out.  If the data
can be partitioned into two clusters, then typically the eigenvector
corresponding to the smallest eigenvalue is a good estimate of the
clustering direction.  The main choice for the user when carrying out
ICS is the choice of the two scatter matrices.

However, in numerical experiments based on a simple mixture of two
bivariate normal distributions, some strange behaviour was noticed.
In certain circumstances, ICS, and its variant PP, badly failed to
pick out the right clustering direction.  Eventually, it was
discovered that the cause was the use of different location measures
in the two scatter matrices.  The purpose of this paper is to explore the
reasons for this strange behaviour in detail and to demonstrate the
benefits of using common location measures.

Section \ref{sec:sec2} gives some examples of scatter matrices and reviews the
use of ICS and PP as clustering methods.  Section \ref{sec:sec3} sets out
the multivariate normal mixture model with two useful standardizations of the 
coordinate system.  Section \ref{sec:sec4} demonstrates in the population setting an ideal situation where
ICS and PP work as expected and where an analytic solution is
available --- the two-group normal mixture
model where the two scatter matrices are given by the covariance
matrix and a kurtosis-based matrix. Some examples with other
robust estimators are given in Sections \ref{sec:sec5}--\ref{sec:sec6}, which show how ICS and PP can go 
wrong when different location measures are used and how the problem is fixed by using a common location measure. 

{\em Notation.}  Univariate random variables, and their realizations,
are denoted by lowercase letters, $x$, say. Multivariate random
vectors, and their realizations, are denoted by lowercase bold
letters, $\boldsymbol{x}$, say. A capital letter, $X$, say is used for
$n \times p$ data matrix containing $p$ variables or measurements on
$n$ observations; $X$ can be written in terms of its rows as
$$
X=\begin{pmatrix} \boldsymbol{x}_1^T\\ \vdots \\ \boldsymbol{x}_n^T 
\end{pmatrix},
$$ 
with $i$th row $\/{x}_i^T=(x_{i1}, \ldots ,x_{ip}),\ i=1,
\ldots, n$.
\section{Background}
\label{sec:sec2}
\subsection{Scatter matrices}
A scatter matrix
$S(X)$, as a function of an $n \times p$ data matrix $X$ is a $p
\times p$ affine equivariant positive definite matrix.  Following
\citet{Tyler}, it is convenient to classify scatter matrices into
three classes depending on their robustness.  
\begin{itemize}
\item [(1)]Class I: is the class of non-robust scatter matrices with
  zero breakdown point and unbounded influence function. Examples
  include the covariance matrix  defined below in \eqref{eq:var} and the
  kurtosis-based matrix  in \eqref{eq:kmat}.
\item [(2)] Class II: is the class of scatter matrices that are
  locally robust, in the sense that they have bounded influence
  function and positive breakdown points not greater than
  $\frac{1}{p+1}$. An example from this class is the class of
  multivariate M-estimators, such as the M-estimate for the
  $t$-distribution  \citep[e.g., ][]{kc94a,arslan1995convergence}.
\item [(3)] Class III: is the class of scatter matrices with high breakdown
  points such as the Stahel-Donoho estimate, the minimum volume
  ellipsoid (mve)  \citep{van2009minimum} and the constrained M-estimates,
  \citep[e.g., ][]{kent1996constrained}.
\end{itemize}
Each scatter matrix has an implicit location measure. Let
us look at the main examples in more detail, and note what happens in
$p=1$ dimension.  The labels in parentheses are used as part of the
notation later in the paper.

The sample covariance matrix (var) is defined by 
\begin{equation}
S = \frac{1}{n} \sum_{i=1}^n (\boldsymbol{x}_i - 
\bar{\boldsymbol{x}})(\boldsymbol{x}_i - \bar{\boldsymbol{x}})^T, 
\label{eq:var}
\end{equation}
where for convenience here a divisor of $1/n$ is used, and where
$\bar{\boldsymbol{x}}$ is the sample mean vector.  The implicit
measure of location is just the sample mean.

The kurtosis-based matrix (kmat) is defined by
\begin{equation}
K = \frac{1}{n} 
\sum_{i=1}^n \{(\boldsymbol{x}_i - \bar{\boldsymbol{x}})^T S^{-1} 
(\boldsymbol{x}_i - \bar{\boldsymbol{x}})\}
(\boldsymbol{x}_i - \bar{\boldsymbol{x}})(\boldsymbol{x}_i - 
\bar{\boldsymbol{x}})^T. \label{eq:kmat}
\end{equation}
Note that outlying observations are given higher weight than for the
covariance matrix, so that $K$ is less robust than $S$.  Again the
implicit measure of location is just the sample mean.  When $p=1$,
the scatter matrix  $S^{-1}K$ reduces to 3 plus the usual univariate kurtosis.

The $M$-estimator of scatter based on the multivariate
$t_\nu$-distribution for fixed $\nu$ is the maximum likelihood
estimate obtained by maximizing the likelihood jointly over scatter
matrix $\Sigma$ and location vector $\/\mu$.  If both parameters are
unknown and  $\nu \geq 1$, then under mild conditions on the data, the mle of
$(\boldsymbol{\mu}, \Sigma)$, is is the unique stationary point of the
likelihood.  Similarly, if  $\nu \geq 0$ and $\/\mu$ is known, the mle of
$\Sigma$, is is the unique stationary point of the likelihood
\citep{kc94a}.  In either case, an iterative
numerical algorithm is needed.  Note that when $\/\mu$ is to be
estimated as well as $\Sigma$, the mle of $\mu$ is the
implicit measure of location for this scatter matrix.  For this paper
we limit attention to the choice $\nu=2$ (and label it below by t2).

The minimum volume ellipsoid (mve) estimate of scatter
$S_{\text{mve}}$, introduced by \citet{rousseeuw1985}, is the
ellipsoid that has the minimum volume among all ellipsoids containing
at least half of observations, and its implicit estimate of location,
$\bar {\boldsymbol{x}}_{\text{mve}}$, say, is the centre of that
ellipsoid. Calculating the exact mve requires extensive computation.
In practice, it is calculated approximately by considering only a
subset of all subsamples that contain $50\%$ of the observations, 
\citep[e.g., ][]{van2009minimum, maronna2006}.  If the location vector is
specified, the search is limited to ellipsoids centred at this
location measure.

When $p=1$, the mve reduces to the lshorth, defined as the length of
the shortest interval that contains at least half of observations.
The corresponding estimate of location,
$\bar{\boldsymbol{x}}_\text{lshorth}$, say, is the midpoint of this
interval. Calculating the lshorth around a known measure of location
is trivial; just find the length of the interval that contains half of
observations centered at this location measure.  The lshorth was
introduced by \cite{grubel1988length}, building on earlier suggestion
of \citet{andrews1972robust} to use
$\bar{\boldsymbol{x}}_\text{lshorth}$, which they called the shorth,
as a location measure.

The minimum covariance determinant estimate of scatter (mcd),
$S_{\text{mcd}}$ is defined as the covariance matrix of half of
observations with the smallest determinant. The mcd location measure,
$\bar{\boldsymbol{x}}_{\text{mcd}}$, say, is the sample mean of those
observations.  The mcd can be calculated approximately by considering
only a subset of all subsamples that contain at least half of
observations, \citep[e.g., ][]{rousseeuw1999fast}.  The mcd estimate of
scatter with respect to a known location measure $\boldsymbol{\mu}$ is
defined as the covariance matrix about $\boldsymbol{\mu}$ of half of
observations with the smallest determinant.  Recall that the
covariance matrix about $\boldsymbol{\mu}$ for a dataset is given by
$S +
(\boldsymbol{\mu}-\bar{\boldsymbol{x}})(\boldsymbol{\mu}-\bar{\boldsymbol{x}})^T$,
where $S$ and $\bar{\boldsymbol{x}}$ are the sample covariance matrix
and mean vector of the dataset.

When $p=1$, the mcd reduces to a truncated variance, $v_{\trunc}$, say, defined as
the smallest variance of half the observations. Its implicit measure
of location, $\bar{x}_{\trunc}$, say, is the sample mean of that interval.  Also, a
modified definition of $v_{\trunc}$ using a known location measure is
trivial and does not require any search; just find the interval that
contains half of observations centered at the given location measure
and calculate the variance.

Routines are available in \texttt{R} \citep{R} to compute (at least
approximately) these robust covariance matrices and their implicit
location measures, in particular, \texttt{tM} from the package
\texttt{ICS} \citep{nordhausen2008} for the multivariate
$t$-distribution, \texttt{cov.rob} from the package \texttt{MASS}
\citep{venables2010package} for mve, and \texttt{CovMcd} from the
package \texttt{rrcov} \citep{todorov2008package} for mcd.  Modified
versions of these routines have been written by us to deal with the
case of known location measures.
\subsection{Invariant coordinate selection and projection pursuit} 
\label{sec:sec22}
Given an $n \times p$ data matrix $X$, the ICS objective function is 
given by the ratio of quadratic forms
\begin{equation}
\kappa_{\text{ICS}}(\boldsymbol{a})=
\frac{\boldsymbol{a}^TS_1\boldsymbol{a}}
{\boldsymbol{a}^TS_2 \boldsymbol{a}}, 
\quad \boldsymbol{a} \in \mathbb{R}^p,
\label{eq:kics}
\end{equation}  
where $S_1=S_1(X)$ and $S_2=S_2(X)$ are two scatter matrices.  By
convention, $S_2$ is chosen to be more robust than $S_1$.  For
exploratory statistical analysis, attention is focused on the choices
for $\boldsymbol{a}$ maximizing or minimizing
$\kappa_{\text{ICS}}(\boldsymbol{a})$.  These values can be calculated
analytically as the eigenvectors of $S_2^{-1}S_1$ corresponding to the
maximum/minimum eigenvalues.

The original ICS method did not make a strong distinction between the
largest and the smallest eigenvalues. However for clustering purposes between
two groups, when the mixing proportion is not too far from $1/2$, it is the
minimum eigenvalue which is of interest; see Section \ref{sec:sec4}.

The method of PP can be regarded as a one-dimensional version of ICS.
It looks for a linear projection $\boldsymbol{a}$ to maximize or
minimize the criterion,
\begin{equation}
\kappa_{\text{PP}}(\boldsymbol{a})=\frac{s_1(X\boldsymbol{a})}
{s_2(X\boldsymbol{a})}. \label{eq:kpp}
\end{equation} 
where $s_1=s_1(X\boldsymbol{a})$ and $s_2=s_2(X\boldsymbol{a})$ are
two one-dimensional measures of spread.  In general, optimizing
$\kappa_{\text{PP}}(\boldsymbol{a})$ must be carried out
numerically. Searching for a global optimum is computationally
expensive, and the complexity of the search increases as the dimension
$p$ increases.  Alternatively, we can search for a local optimum
starting from a sensible initial solution, such as the ICS optimum
direction.

Both ICS and PP are equivariant under affine transformations.  That
is, if $X$ is transformed to $U= \boldsymbol{1}_n \boldsymbol{h}^T +
XQ^T$, where $Q(p \times p)$ is nonsingular and $\boldsymbol{h}$ is a
translation vector in $\mathbb{R}^p$, then for either ICS or PP the
new optimal vector $\boldsymbol{b}$, say, for $U$ is related to
  the corresponding optimal vector $\boldsymbol{a}$ for $X$ by
\begin{equation}
\boldsymbol{b} \propto Q^{-T}\boldsymbol{a}. \label{eq:scale}
\end{equation}

For numerical work it is convenient to have an explicit notation for the 
different choices in ICS and PP.  If Scat1 and Scat2 are the names of
two types of multivariate scatter matrix, each computed with its own implicit
location measure, then the corresponding versions of ICS and PP will be
denoted
\begin{equation}
\text{ICS}:\text{Scat1}:\text{Scat2}, \quad \text{and} \quad
\text{PP}:\text{Scat1}:\text{Scat2}. \nonumber 
\end{equation}
Note that PP is based on the univariate versions of Scat1 and Scat2.
For example, ICS based on the covariance matrix and the minimum volume
ellipsoid will be denoted by ICS:var:mve.  Other choices for scatter
matrices have been summarized in Section \ref{sec:sec2}.

 When a common location measure is imposed on Scat1 and Scat2, then
 this restriction will be indicated by the augmented notation
 \begin{equation}
 \text{ICS}:\text{Scat1}:\text{Scat2}:\text{Loc}, \nonumber  
 \end{equation}
 and similarly for PP.  In this paper the only choice used for the
 location measure is the sample mean (mean).  For example, ICS based on the
 covariance matrix and the minimum volume ellipsoid, both computed
 with respect to the mean vector, is denoted
     $$
     \text{ICS}:\text{var}:\text{mve}:\text{mean}.
     $$
\section{The two-group multivariate normal mixture model}
\label{sec:sec3}
The simple model used to demonstrate the main points of this paper is the
two group multivariate normal mixture model, with density
\begin{equation}
f(\boldsymbol{x}) = q \phi_p(\boldsymbol{x}, \boldsymbol{\mu}_1, \Omega) +
 (1-q) \phi_p(\boldsymbol{x}, \boldsymbol{\mu_2}, \Omega), \nonumber
\end{equation}
where $\phi_p$ is the multivariate normal density, $\boldsymbol{\mu}_1$ and $\boldsymbol{\mu}_2$
are two mean vectors, $\Omega$ is a common covariance matrix, and
$0<q<1$ is the mixing proportion.  Even in this simple case, major 
problems with ICS and PP can arise.

Since ICS and PP are affine equivariant, we may without loss of
generality choose the coordinate system so that 
$$
\boldsymbol{\mu_1} = \alpha \boldsymbol{e_1}, \quad \boldsymbol{\mu_2} = -\alpha \boldsymbol{e_1}, \quad \Omega = I_p,
$$
where $\boldsymbol{e_1} = (1,0,\ldots, 0)^T$ is a unit vector along
the first coordinate axis, and $\alpha>0$.  That is,
$\boldsymbol{\mu_1}$ and $\boldsymbol{\mu_2}$ lie equally spaced about
the origin along the first coordinate axis, and the covariance matrix
of each component equals the identity matrix.

A random vector $\/x$ from the mixture model can also be given a 
stochastic representation, 
\begin{equation}
\boldsymbol{x} = \alpha s \boldsymbol{e}_1 + \boldsymbol{\epsilon}, \nonumber
\end{equation}
where $\boldsymbol{\epsilon} \sim N_p(0,I_p)$ independently of an indicator variable $s$, 
\begin{equation}
s=\left\{
\begin{array}{rl}
1& \text{ with probability } q\\
-1& \text{ with probability } (1-q)\\
\end{array} \right. \nonumber.  
\end{equation}
Moments under the mixture model are calculated most simply in terms of this
stochastic representation.  In particular, 
\begin{equation}
\boldsymbol{\mu}_x= E(\boldsymbol{x}) = q \boldsymbol{\mu}_1+ (1-q) \boldsymbol{\mu}_2=(2q-1)\alpha \boldsymbol{e}_1, \quad
E(\boldsymbol{xx}^T) = \alpha^2 \boldsymbol{e}_1\boldsymbol{e}_1^T + I_p, \nonumber
\end{equation} 
so that the  covariance matrix is 
\begin{equation}
\label{eq:sigx}
\Sigma_x=\var(\boldsymbol{x}) = E(\boldsymbol{xx}^T) - \boldsymbol{\mu}_x \boldsymbol{\mu}_x^T = 
4q(1-q)\alpha^2 \boldsymbol{e}_1 \boldsymbol{e}_1^T + I_p.
\end{equation}
For practical work it is also convenient to consider a 
standardization for which the overall covariance matrix is the
identity matrix.  That is, define a new random vector
\begin{equation}
\boldsymbol{y} = C^{-1}\boldsymbol{x}, \label{eq:y} 
\end{equation}
where $C^{-1} = \diag(1/c_1,\ldots, 1/c_p)$, where $c_1 = \{1+4
q(1-q)\alpha^2\}^{1/2}$, and $c_2 = \cdots = c_p = 1$.  Then $\boldsymbol{y}$
has a stochastic representation
$$
\boldsymbol{y} = \delta s \boldsymbol{e}_1 + \boldsymbol{\eta},
$$
where 
\begin{equation}
\label{eq:delta}
\delta = \alpha / \{1+4q(1-q)\alpha^2 \}^{1/2},
\end{equation}
and
\begin{equation}
\boldsymbol{\eta }\sim N_p(0, \diag(\sigma^2_\eta, 1, \ldots, 1)) \nonumber 
\end{equation}
where the first diagonal term $\sigma^2_\eta$ has two equivalent formulas,
$$
\sigma^2_\eta=\{1+4 \alpha^2 q(1-q)\}^{-1} \quad \text{or} \quad 
\sigma^2_\eta=1-4q(1-q)\delta^2
$$

The first two moments of $\boldsymbol{y}$ are
$$
\boldsymbol{\mu}_y = (2q-1)\delta \boldsymbol{e}_1, \quad \Sigma_y = I_p.
$$
\section{A population example: PP based on the kurtosis and ICS based on the kurtosis-based matrix and the covariance matrix}
\label{sec:sec4}
In this section we look at ICS:kmat:var and PP:kmat:var in the
population case.  In this setting it is possible
to derive analytic results.  Note that since kmat is based on fourth
moments it is less robust than the variance matrix; hence kmat is
listed first. 

Recall the kurtosis of a univariate random variable $u$, say, with mean
$\mu_u$, is defined by
\begin{equation}
\text{kurt}(u)=\frac{\text{E}\{(u-\mu_u)^4\}}{\left[\text{E}\{(u-\mu_u)^2\}\right]^2}-3. \nonumber
\end{equation} 
The univariate kurtosis is zero when the random variable has normal
distribution. For non-normal distributions the kurtosis lies in the
interval $[-2, \infty]$ and is often nonzero.

\citet{pena2001} studied the population version of PP:kmat:var and
showed that when the mixing proportion is not too far from $1/2$ (more
precisely, if $q(1-q)>1/6$, i.e. $0.21 < q <0.79$), then minimizing the
PP objective function picks out the correct clustering direction.

Their result can be derived simply as follows.  Let $\/a \in
\mathbb{R}^p$ be a unit vector.  Write $\/a^T\/x = \alpha a_1 s  + v$,
where $v = \/{a}^T \/{\epsilon} \sim N(0,1)$ is independent of $s$.  
The moments of $s$ are E$(s)=$E$(s^3)=m$, say, where
\begin{equation}
\label{eq:c}
m= 2q-1,
\end{equation}
and E$(s^2)=$ E$(s^4) = 1$. Hence, var$(s)= \sigma^2$, say, where
\begin{equation}
\label{eq:sigma}
\sigma^2= 4q(1-q).
\end{equation}
Then 
$$
\text{kurt}(s) =  - 6 + 4/\sigma^2.
$$
It can be checked that $\text{kurt}(s)<0$ provided $q(1-q)>1/6$.

Next, we use the property that if $u_1,  u_2$ are independent random
variables with the same variance, and if $\delta_1, \delta_2$ are
coefficients satisfying $\delta_1^2 + \delta_2^2 = 1$, then
$$
\kurt(\delta_1 u_1 + \delta_2 u_2 ) = 
\delta_1^4 \kurt(u_1) + \delta_2^4 \kurt(u_2).
$$
Applying this result to $\/a^T\/x$ yields
\begin{equation}
\label{eq:PP}
\kurt(\/a^T\/x) =\frac{a_1^4 \alpha^4 \sigma^4}{(\alpha^2 a_1^2 \sigma^2+1)^2}
\kurt(s). 
\end{equation}
Provided $\kurt(s)<0$, \eqref{eq:PP} is minimized when $a_1^2$ is
maximized, that is, if $a_1^2=1$, so that $\/a = \pm e_1$ picks out the
first coordinate axis.

The ICS calculations proceed similarly.  First note that, the first diagonal term in
 $\Sigma_x$, defined in \eqref{eq:sigx}, can be expressed in terms of $\sigma^2$, defined in \eqref{eq:sigma}, as
  $\alpha^2 \sigma^2 + 1$.

The first factor in the population version of $K$ defined in \eqref{eq:kmat}, $K_x$, say, is given by
\begin{equation}
(\boldsymbol{x}-\boldsymbol{\mu}_x)^T \Sigma_{\boldsymbol{x}}^{-1}(\boldsymbol{x}-\boldsymbol{\mu}_x)= 
\frac{(x_1-\alpha m)^2}{1+\alpha^2 \sigma^2}+x_2^2 + \cdots + x_p^2 = D^2, \text{ say,}
\nonumber
\end{equation}
where $m$ is defined in \eqref{eq:c}.
Note that $D^2$ is an even function in $x_2, \ldots, x_p$.  Hence by symmetry
all the off-diagonal terms in $K_x$ vanish.  The first diagonal term is
given by 
\begin{equation}
E\{D^2 (x_1 - \alpha m)^2\} = (1+\alpha^2 \sigma^2)(p+2)+
  \frac{\alpha^4 \sigma^4 \text{kurt}(s)}{(1+\alpha^2 \sigma^2)}. \nonumber
\end{equation}
The remaining diagonal terms, $j=2, \ldots, p$ are given by
\begin{equation}
E\{D^2 x_j^2\}  = p+2. \nonumber
\end{equation}
Hence $\Sigma_x^{-1} K_x$ reduces to 
$$
\diag(p+2+\frac{\text{kurt}(s) \alpha^4 \sigma^4}{(1+\alpha^2 \sigma^2)}, p+2, \ldots, p+2).
$$
These diagonal values are the eigenvalues.  Hence provided $\kurt(s)<0$,
$\kappa_{\text{ICS}}$ is minimized when $\boldsymbol{a} = \boldsymbol{e}_1$, that is, when $\boldsymbol{a}$ picks out the clustering
direction.

\begin{figure}[h!] 
\centering
\includegraphics[width=6cm,height=6cm]{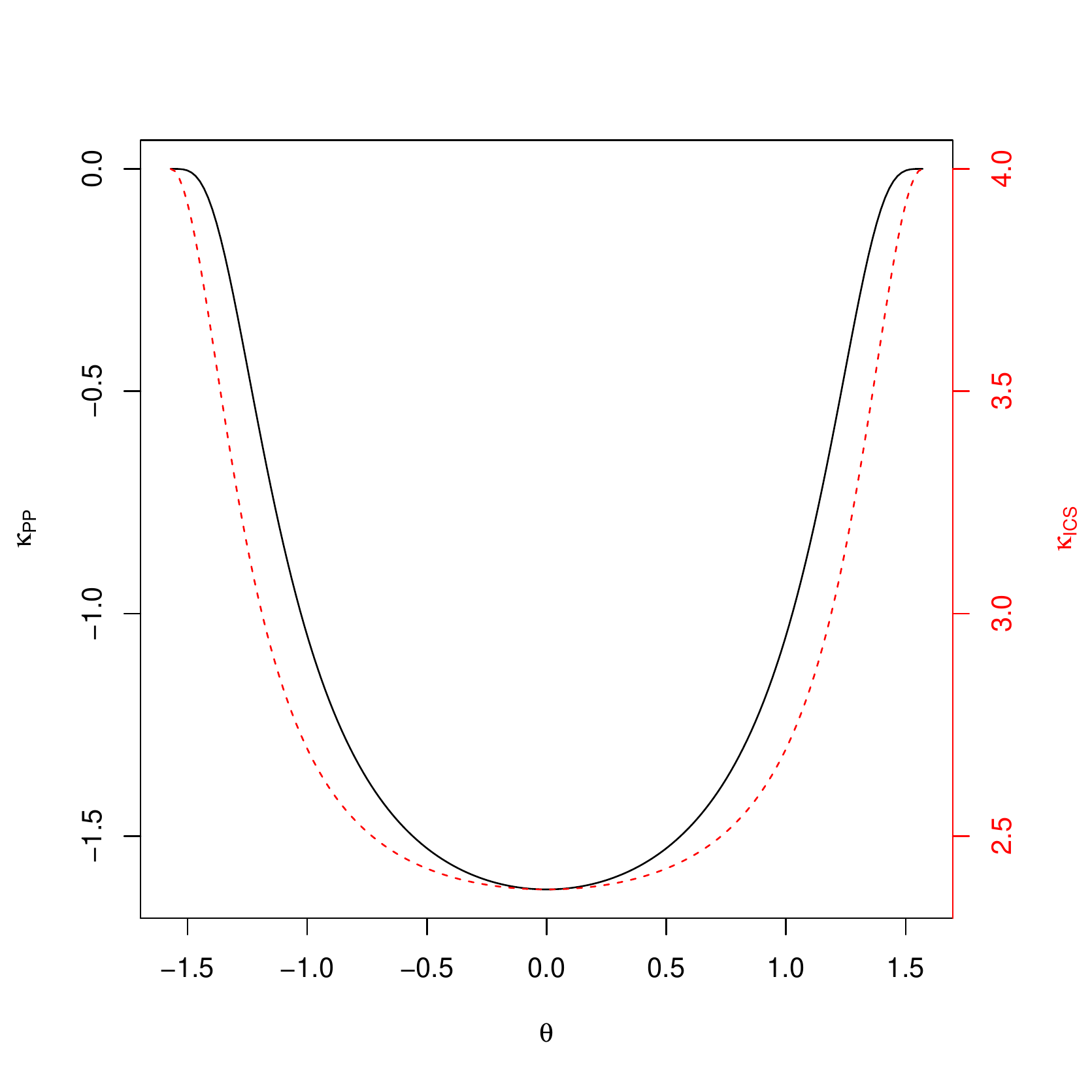}
\caption{Plot of the population criteria $\kappa_{\text{ICS}}(\theta)$ (red dotted line), and $\kappa_{\text{PP}}(\theta)$ (solid black line) versus $\theta$, for $q=1/2$, $\alpha=3$.}
\label{fig:fig1}
\end{figure}

If $p=2$, we can write a unit vector as $\boldsymbol{a} =
(\cos \theta, \sin \theta)^T$, and since $\/a$ and $-\boldsymbol{a}$
define the same axis, we can parameterize the ICS and PP objective
functions in terms of $\theta, \ -\pi/2 \leq \theta \leq \pi/2$.
Plots of $\kappa_\ICS(\theta)$ and $\kappa_\PP(\theta)$ for $\alpha=3$
and $q=1/2$ are shown in Figure \ref{fig:fig1}.

For numerical work, especially when the underlying mixture model is
unknown, the only feasible standardization is to ensure the overall
variance matrix $\Sigma_y$ is the identity rather than the within
group variance matrix.  In terms of the population model of this
section, it means working with $\/y$ from \eqref{eq:y} rather than
$\/x$.  If $p=2$ and $\/b \propto (\cos \phi, \sin \phi)^T$, say, is
also written in polar coordinates, then from \eqref{eq:scale} and
\eqref{eq:y} $\/a$ and $\/b$ are related by
$$
\/b \propto  C \/a, 
$$
hence, $\phi$ and $\theta$ are related by
\begin{equation}
\begin{pmatrix} \cos \phi \\ \sin \phi \end{pmatrix} \propto  \begin{pmatrix} c_1 & 0 \\ 0& c_2 \end{pmatrix} 
\begin{pmatrix} \cos \theta \\ \sin \theta \end{pmatrix}. \nonumber
\end{equation}
Thus,
$$
\tan \phi = c \tan \theta,
$$
where $c=c_2/c_1$. 

\begin{figure}[h!] 
\centering
\includegraphics[width=6cm,height=6cm]{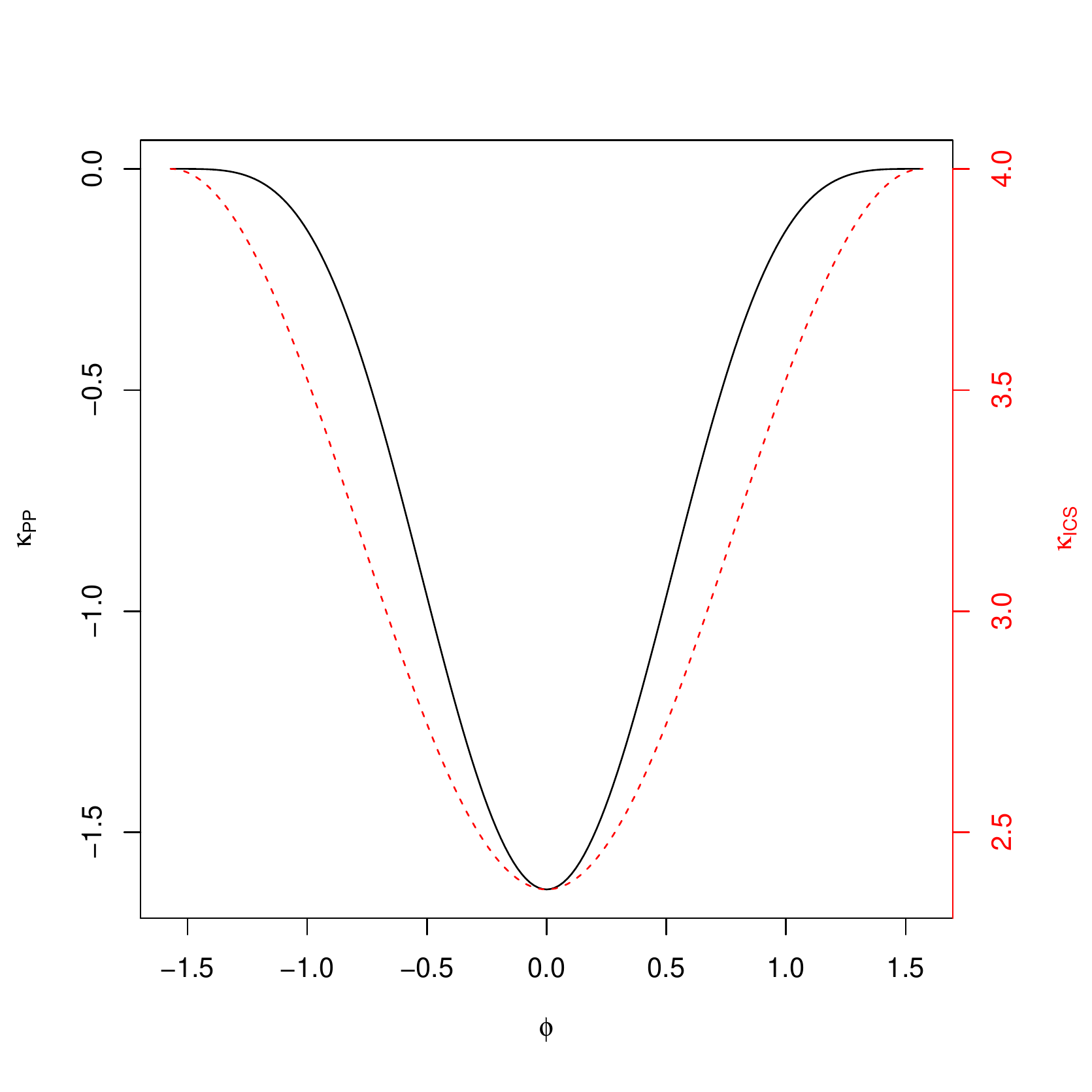}
\caption{Plot of the population criteria $\kappa_{\text{ICS}}(\phi)$ (red dotted line), and $\kappa_{\text{PP}}(\phi)$ (solid black line) versus $\phi$, for $q=1/2$, and $\delta=0.95$.}
\label{fig:fig2}
\end{figure}

The plot of the ICS and PP objective functions in Figure
\ref{fig:fig2} shows that there is a sharper minimum in $\phi$ coordinates than in  $\theta$ coordinates because 
under our mixture model $c$ is less than 1. If $\/x$ is scaled as in \eqref{eq:y} with $c_1>c_2$, i.e $c>1$, then there
 will be a wider minimum in $\phi$.   
\section{The effect of using a common location measure 
on ICS and PP} 
\label{sec:sec5}
As we mentioned earlier in Section \ref{sec:sec22}, the ICS and PP
criteria are expected to have
similar behaviour to the kurtosis-based criteria in Section
\ref{sec:sec4}. Namely, they are expected to be minimized in the
clustering direction when the mixing proportion is not too far from 1/2.

However, when applying ICS with at least one robust estimate of
scatter (mainly from Class III), some peculiar behaviour was observed.
In particular, the ICS criterion was often  maximized in the clustering
direction rather than  minimized.

Here is an  explanation.  Under the two-group mixture model with one
group slightly bigger than the other, a class III scatter matrix will
typically home in on the larger group, with its corresponding location
measure at the center of this group and its estimate of the scatter
matrix capturing the spread of this group. The other scatter matrix
(Class I or II) will measure the overall scatter of the data with its
corresponding location measure at the overall center of the data.  The
result is erratic behaviour in $\kappa_\ICS$ and $\kappa_\PP$.

Imposing a common location measure on the two scatter matrices fixes
this problem.  Here is an example in $p=2$ dimensions to
illustrate the issues in greater detail.

In this example we look at ICS:var:mve  for the
population bivariate normal mixture model in Section \ref{sec:sec3},
with $q=1/2$ and any value of $\alpha>0$, i.e. $0 \leq \delta \leq 1$, where $\delta$ is given in \eqref{eq:delta}.
Standardize the coordinate
system so that the overall covariance matrix is the identity,
$\Sigma_y=I_2$. Let $\Sigma_{\text{mve}}$ denote the population
minimum volume ellipsoid scatter matrix.

Then it turns out that  $\Sigma_{\text{mve}}$ is the within-group
covariance matrix for (either) one of the groups,
\begin{equation}
\label{eq:mvenc}
\Sigma_{\text{mve}}=\begin{pmatrix}1-\delta^2 &0 \\ 0&1  \end{pmatrix}, 
\end{equation}   
where $0 \leq \delta \leq 1$ is given in (\ref{eq:delta}).  The
implicit estimate of the center of the data will be given by the
center of either group, $\pm \delta \boldsymbol{e}_1$; both values fit equally
well.

\begin{figure}[h!] 
\centering
\includegraphics[width=6cm,height=6cm]{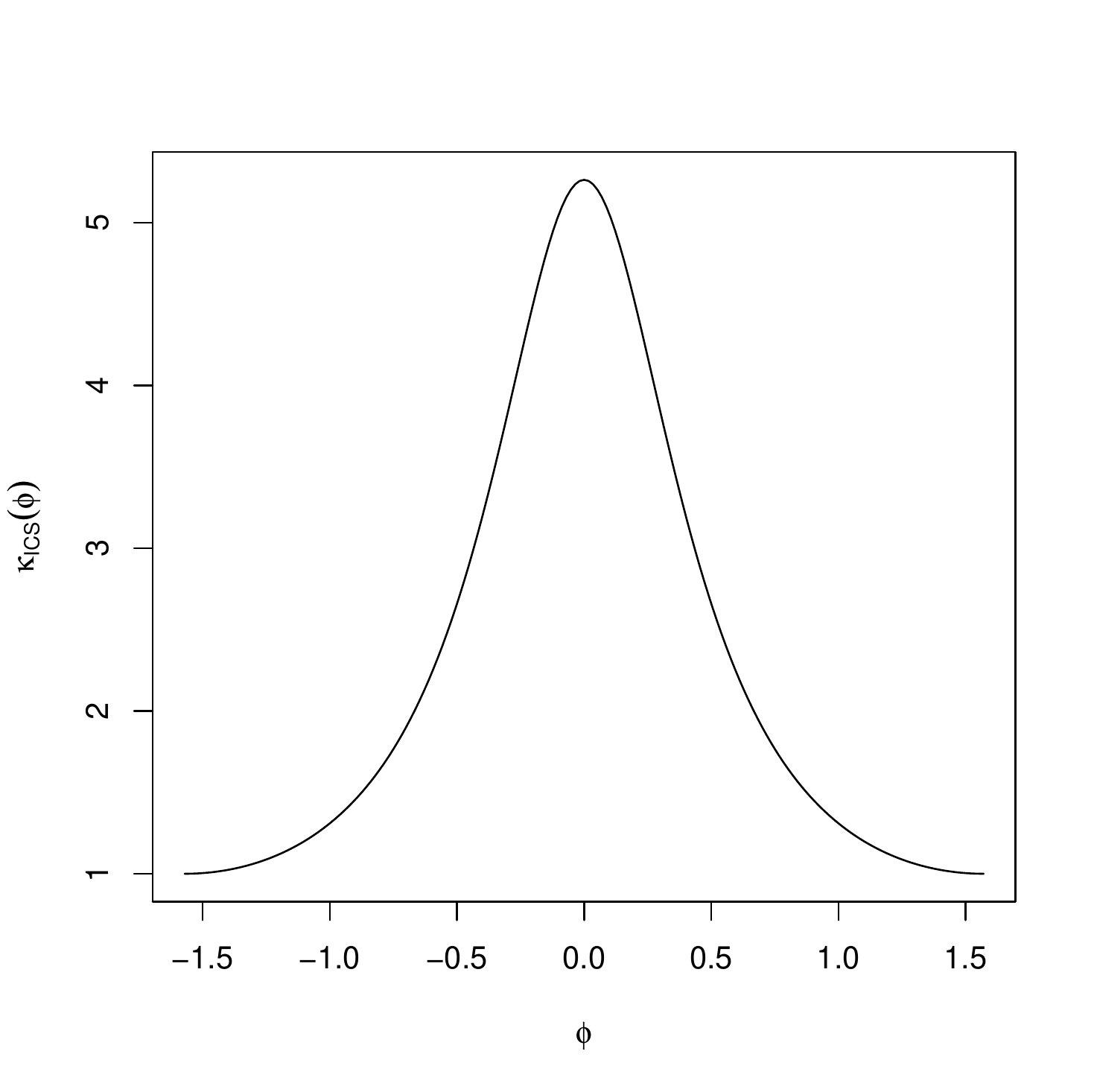}
\caption{Plot of the population criterion of ICS:var:mve vs. $\phi$ for $\delta=0.9$.}
\label{fig:fig3}
\end{figure}

Figure \ref{fig:fig3} shows that the clustering direction estimated by the ICS:var:mve method is the direction
that minimizes $\kappa_{\text{ICS}}$ (the eigenvector of the smallest eigenvalue of $\Sigma_{\text{mve}}^{-1}$),
 namely $(0,1)^T$, i.e. $\phi=\pm \pi/2$. However, the true direction of group separation 
direction is $(1,0)^T$, i.e. $\phi=0$.

Next consider ICS:var:mve:mean, i.e. the common mean version of the previous example.  The overall mean of the data 
is at
the origin.  When $\Sigma_{\text{mve}}$ is constrained to have its
location measure at the origin, then the ICS criterion now picks out the true
clustering direction.  In order to give an analytic proof of this result,
we restrict attention to the the limiting case of the balanced mixture model, i.e 
when $\delta=1,\ q=1/2$.
Hence, the group components will lie on two parallel vertical lines with means
\begin{equation}
\mu_1=(1,0)^T, \ \ \mu_2=(-1,0)^T, \nonumber
\end{equation}
and within-group covariance matrix  
\begin{equation}
\begin{pmatrix}0 & 0 \\0 & 1 \end{pmatrix}. \nonumber
\end{equation}

\begin{figure}[h!] 
\centering
\includegraphics[width=9cm,height=9cm]{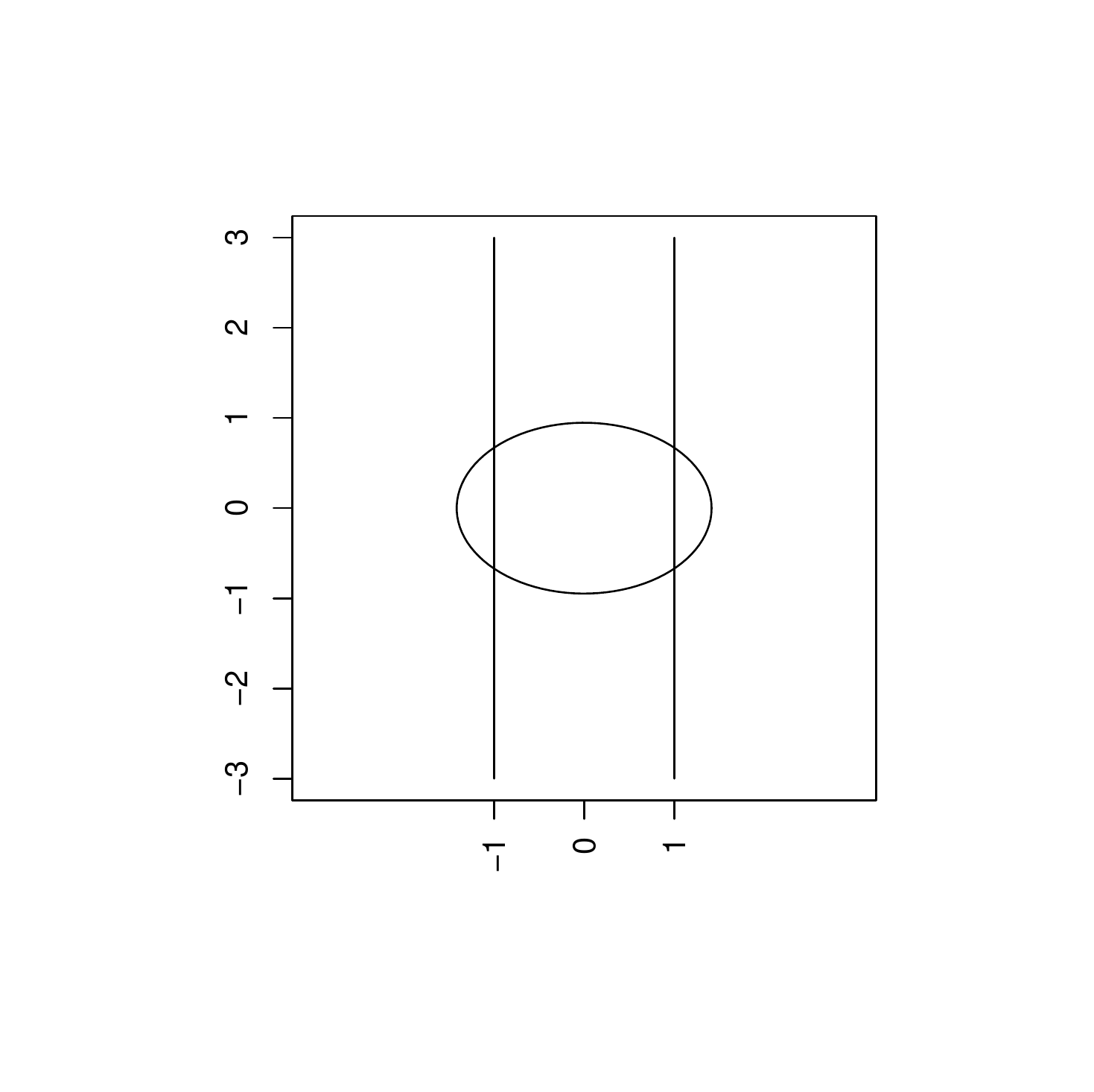}
\caption{Plot of the ellipse of the constrained $\Sigma_{\text{mve}}$.}
\label{fig:mve}
\end{figure}

\begin{figure}[h!] 
\centering
\includegraphics[width=6cm,height=6cm]{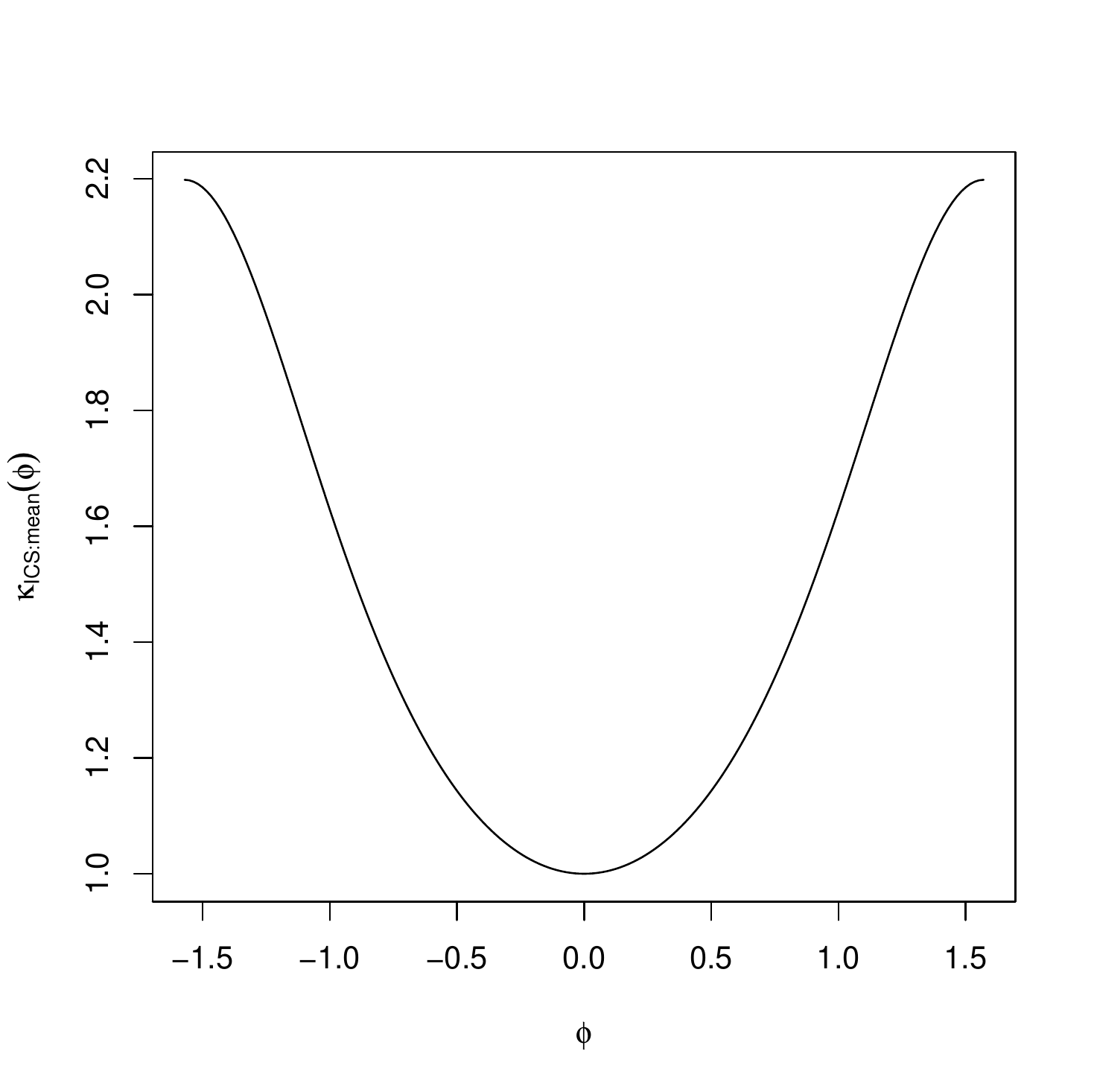}
\caption{Plot of the population criteria of ICS:var:mve:mean, $\kappa_{\text{ICS}:\mu}(\phi)$.}
\label{fig:fig4}
\end{figure}

In this setting, it can be shown that the population version of the MVE matrix,
$\Sigma_{\text{mve}}$, say, takes the form
\begin{equation}
\Sigma_{\text{mve}}=c_t \Sigma_t= \begin{pmatrix}2 &0 \\ 0& 2d^2 \end{pmatrix},
 \nonumber
\end{equation}
where $d = \Phi^{-1}(.75) = 0.674$, the 75th quantile of the standard
normal distribution (see the Appendix).  Hence the dominant
eigenvector is $\/e_1$.  The ellipse of $\Sigma_{\text{mve}}$ is
plotted in Figure \ref{fig:mve}.  Figure \ref{fig:fig4} shows that the
criterion of ICS:var:mve:mean, $\kappa_{\ICS:\mu}(\phi)$ picks out the
correct clustering direction $\/e_1$.


Like ICS, PP can fail to detect the clustering direction if applied
using different location measures.  The reason for that is the
projection direction that separates the data into two groups with one
slightly bigger than the other, the more robust measure of spread will
be located at the larger group.  In Section \ref{sec:sec6}, we give a
detailed numerical example of the problem arising from using two
different location measures in PP:var:mcd, and how the problem is
fixed by using a common location measure.
\section{Examples} 
\label{sec:sec6}
\subsection*{Overview}
In this section, we give numerical examples that demonstrate different
ways in which ICS and/or PP can go wrong.  We also show the effect of
using common location measures in these examples.  We use one
simulated data set and apply different ICS and PP methods, with and
without imposing a common location measure (the mean).

A two-dimensional data set of size $n=500$ is generated from the
balanced mixture model, defined in Section \ref{sec:sec3}, with
$q=1/2$, and $\alpha=3$, so that $\delta=0.95$.  Thus the two groups
are well-separated and no sensible statistical method should have any
problem finding the two clusters.  All calculations are done after
standardization with respect to the ``total'' coordinates.  That is,
the data matrix $Y (500 \times 2)$ is standardized to have sample mean
$\/0$ and sample covariance matrix $I_2$.

The ICS and PP methods used are:
\begin{itemize}
\item[(1)] (PP,ICS):var:t2 with corresponding criteria $\kappa_{\text{ICS}}^{1}$, 
and $\kappa_{\text{ICS}}^{1}$.
\item[(2)](PP,ICS):var:mcd with corresponding criteria $\kappa_{\text{ICS}}^{2}$, and $\kappa_{\text{PP}}^{2}$.
\item[(3)](PP,ICS):var:mve with corresponding criteria $\kappa_{\text{ICS}}^{3}$, and $\kappa_{\text{PP}}^{3}$.
\item[(4)](PP,ICS):t2:mcd with corresponding criteria $\kappa_{\text{ICS}}^{4}$, and $\kappa_{\text{PP}}^{4}$.
\item[(5)](PP,ICS):t2:mve with corresponding criteria $\kappa_{\text{ICS}}^{5}$, and $\kappa_{\text{PP}}^{5}$.
\end{itemize} 
When imposing the mean as the common location measure, the ICS and PP criteria will be denoted by $\kappa_{\text{ICS:mean}}^{j}$
 and $\kappa_{\text{PP:mean}}^{j}$, where $j=1, \ldots, 5$.
 
To understand the behaviour of the ICS and PP, their criteria are plotted against $-\pi/2\leq\phi\leq\pi/2$. 
The plots are shown in Figure \ref{fig:fig5}.
\newpage
\begin{figure}[t!]
\centering
\subfloat[(PP,ICS):var:t2]{%
\includegraphics[width=5cm,height=5cm]{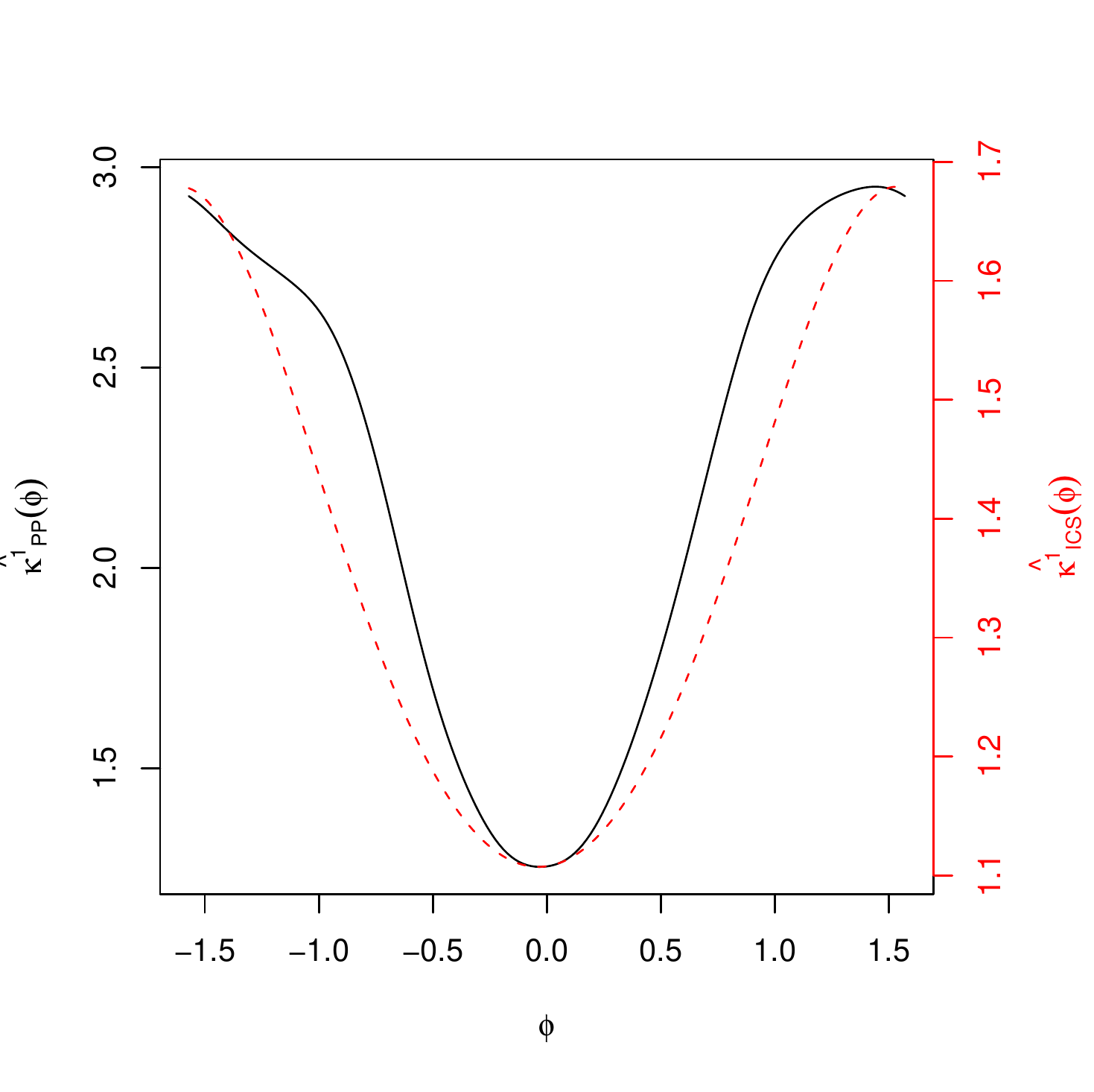}%
}%
\subfloat[(PP,ICS):var:t2:var]{%
\includegraphics[width=5cm,height=5cm]{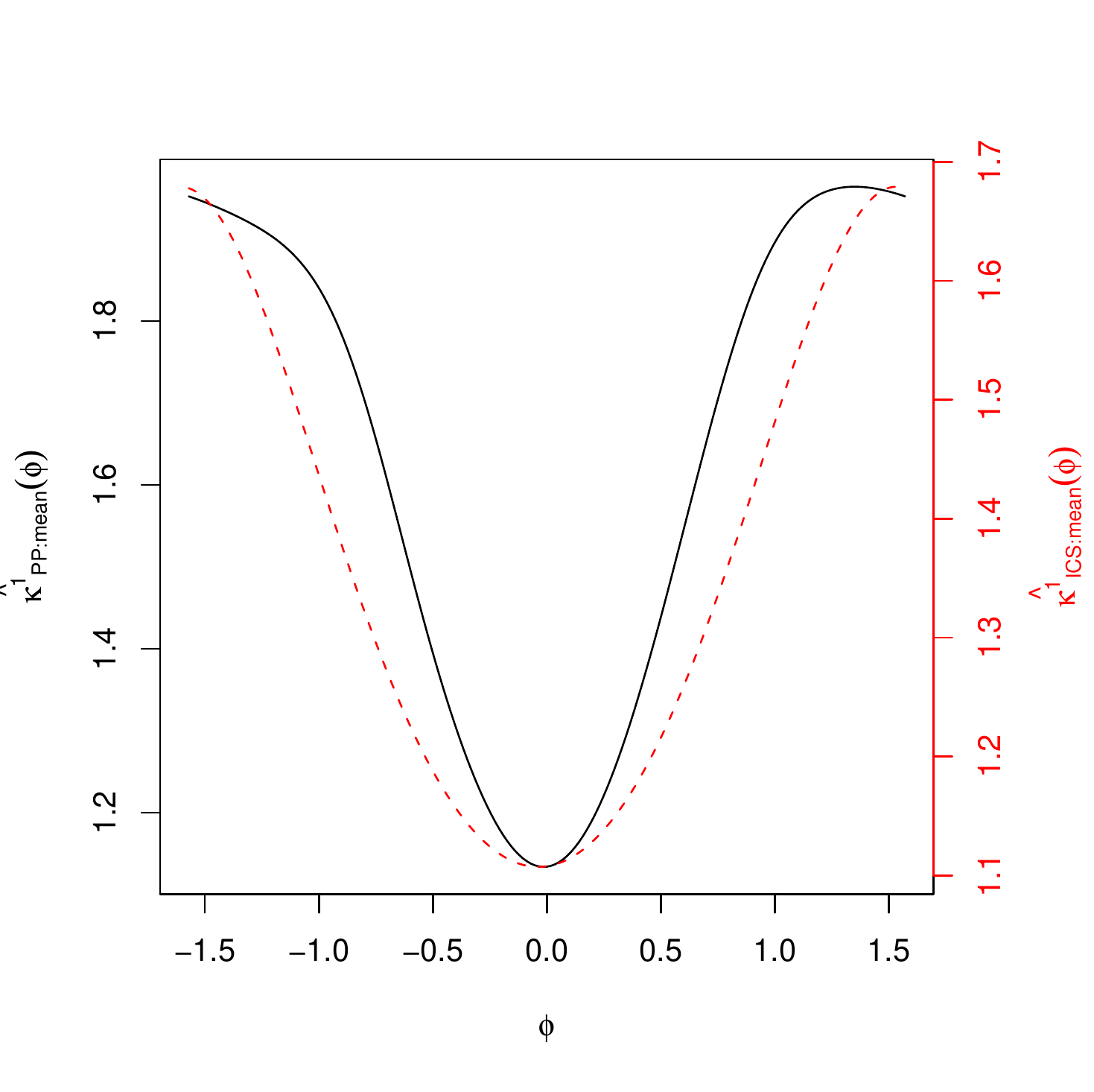}}
\\
\centering
\subfloat[(PP,ICS):var:mcd]{\includegraphics[width=5cm,height=5cm]{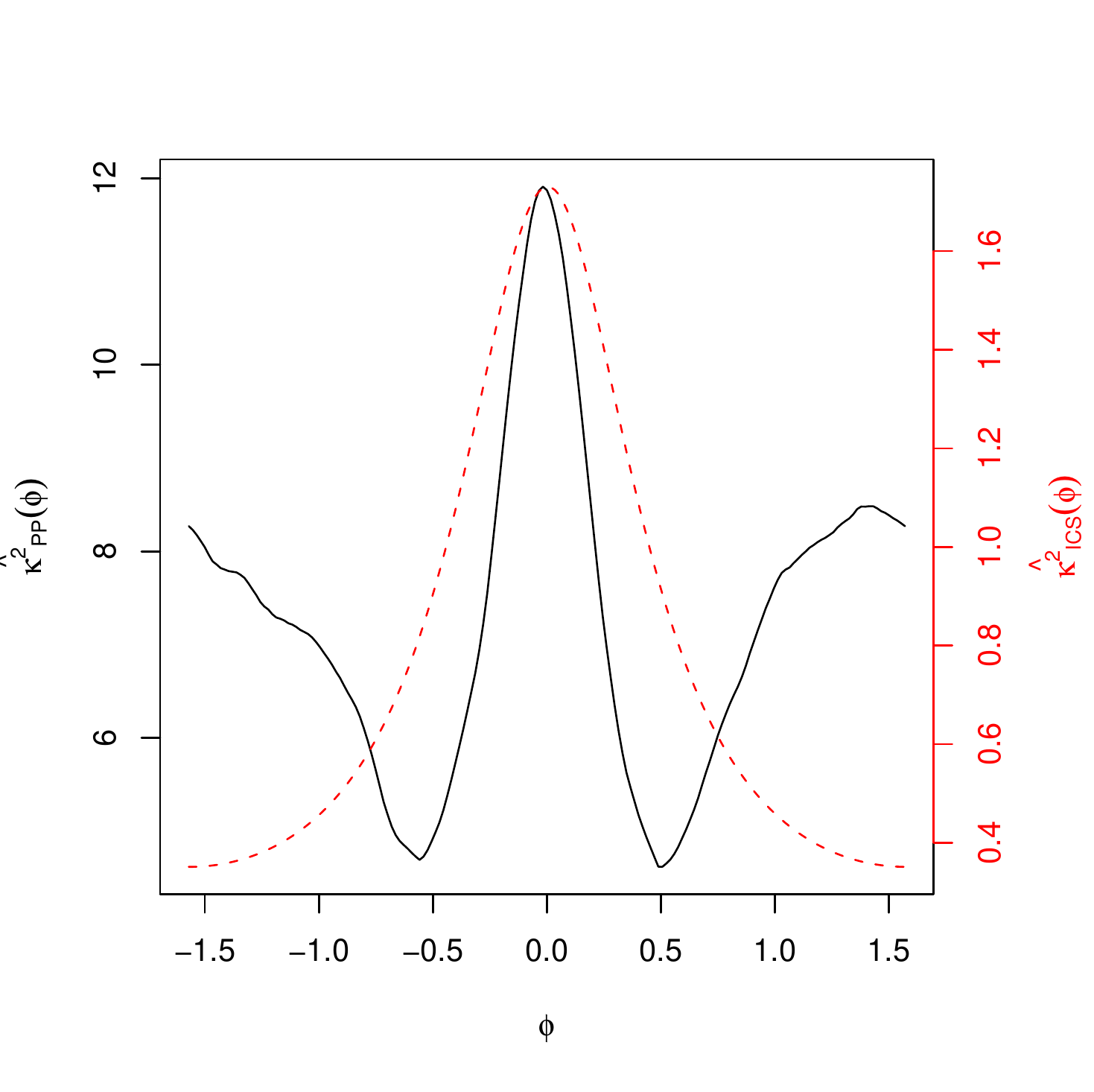}}%
\subfloat[(PP,ICS):var:mcd:mean]{\includegraphics[width=5cm,height=5cm]{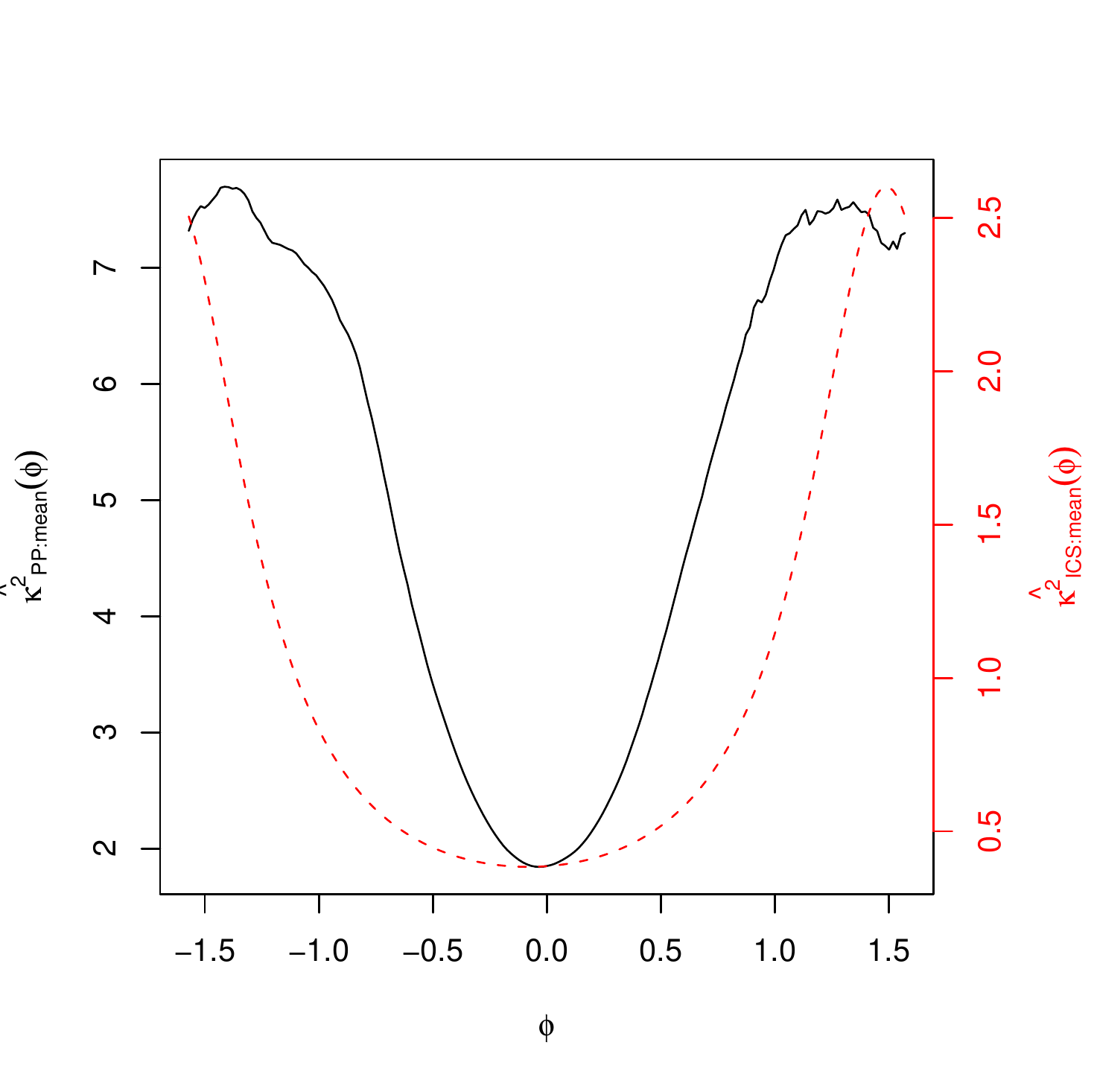}}
\\
\centering
\subfloat[(PP,ICS):var:mve]{\includegraphics[width=5cm,height=5cm]{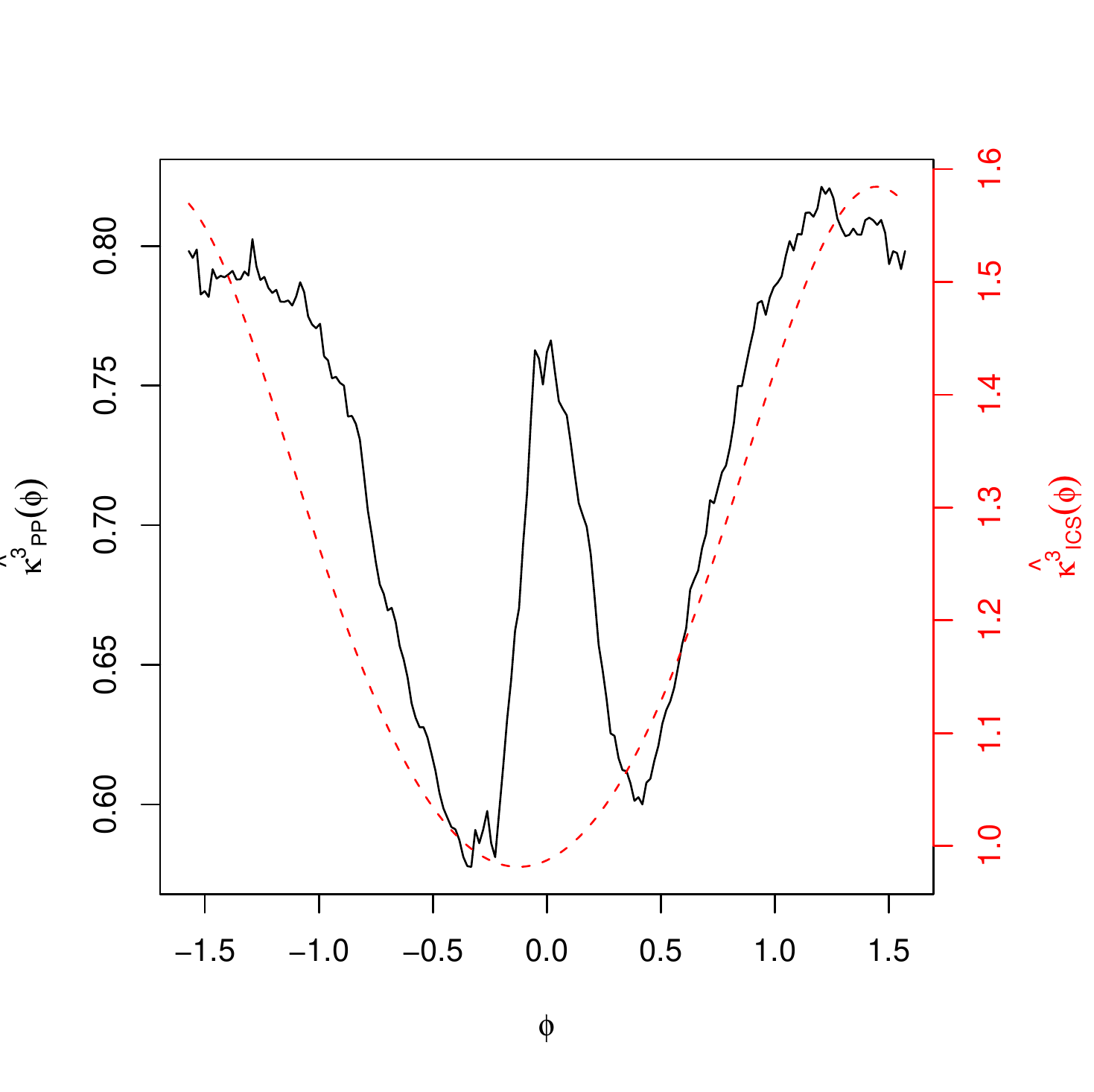}}%
\subfloat[(PP,ICS):var:mve:mean]{\includegraphics[width=5cm,height=5cm]{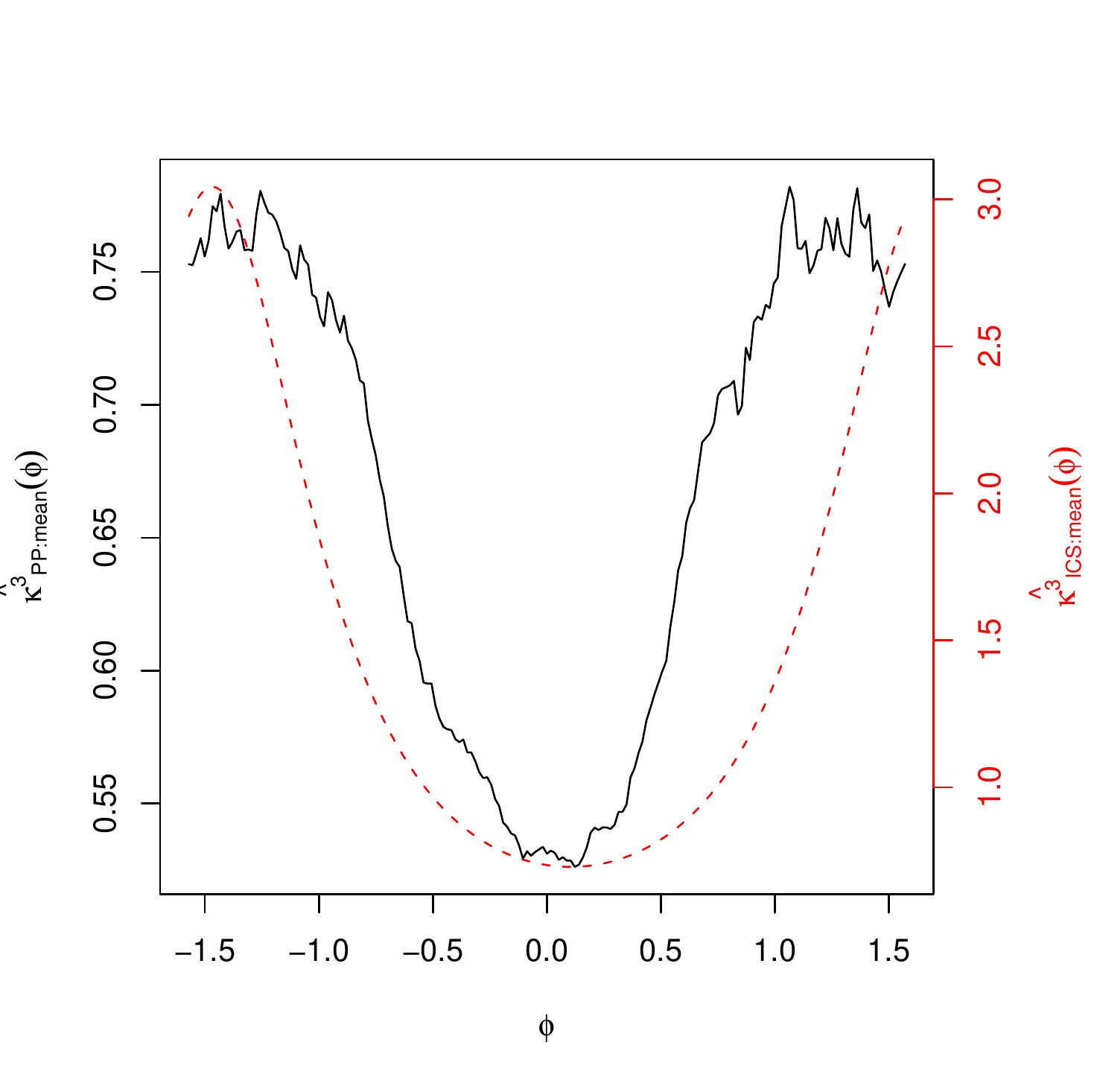}}
\caption{}
\end{figure}

\newpage
\begin{figure}
\ContinuedFloat
\centering
\subfloat[(PP,ICS):t2:mcd]{\includegraphics[width=5cm,height=5cm]{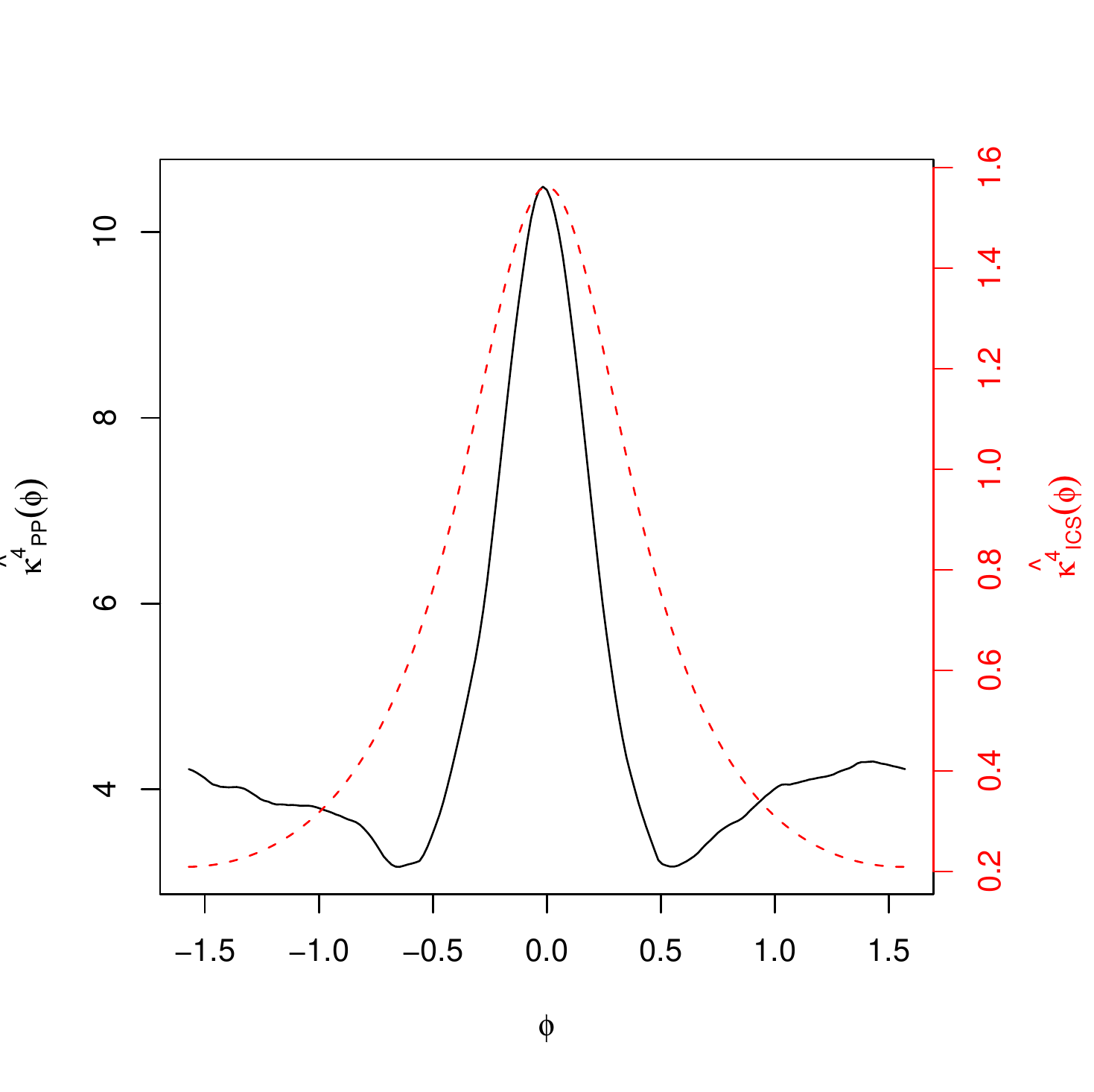}}%
\subfloat[(PP,ICS):t2:mcd:mean]{\includegraphics[width=5cm,height=5cm]{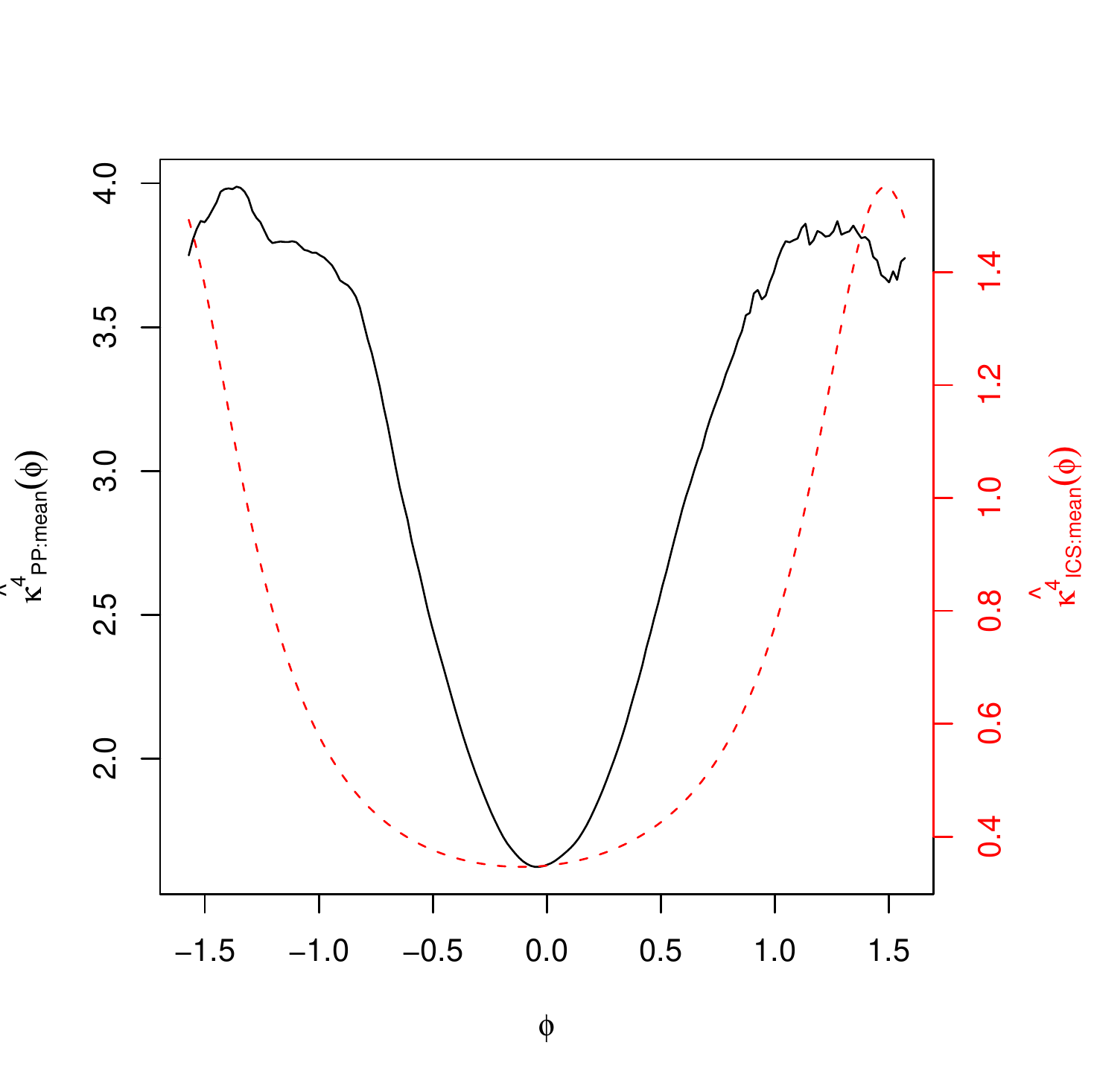}}
\\
\centering
\subfloat[(PP,ICS):t2:mve]{\includegraphics[width=5cm,height=5cm]{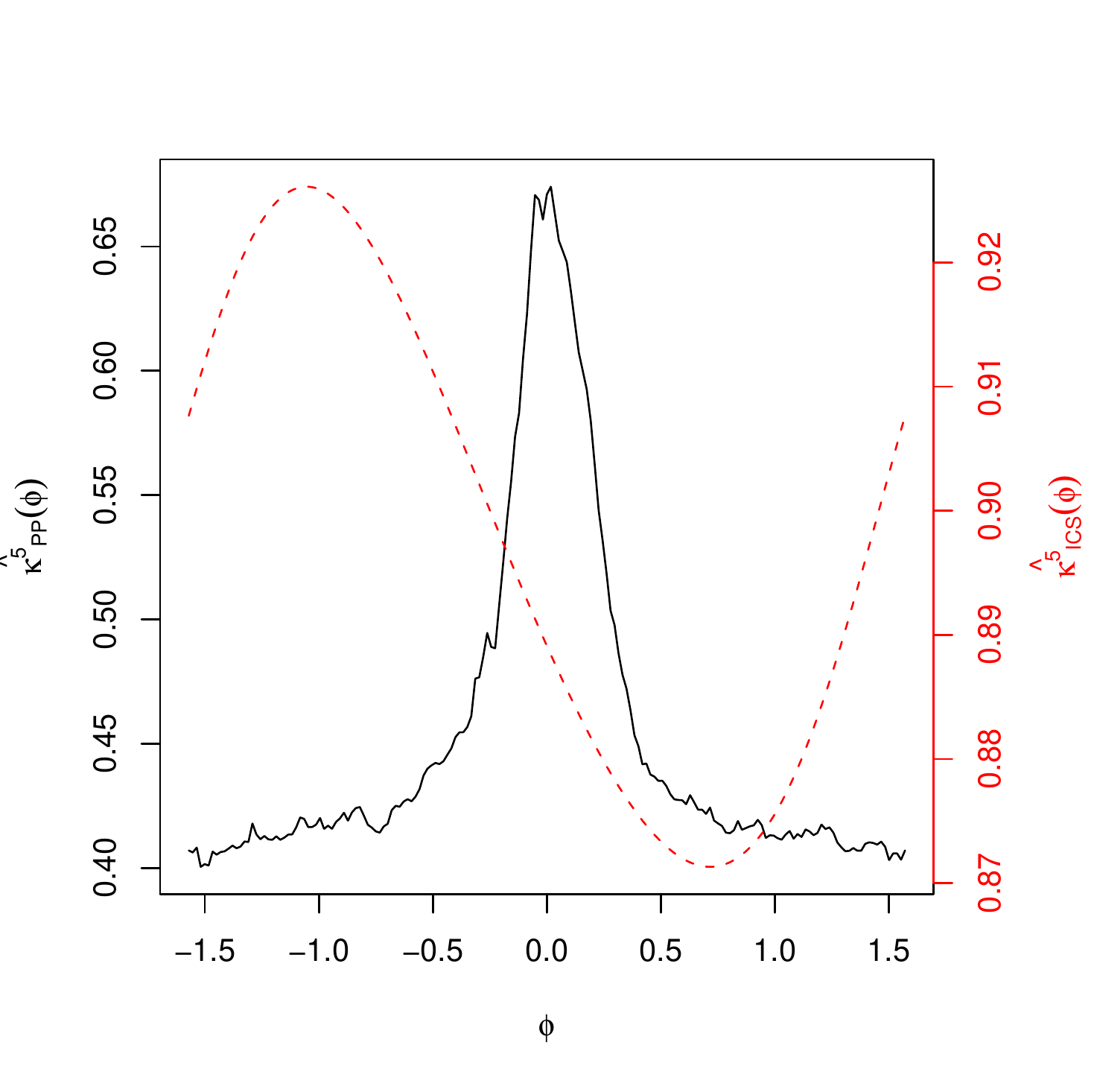}}%
\subfloat[(PP,ICS):t2:mve:mean]{\includegraphics[width=5cm,height=5cm]{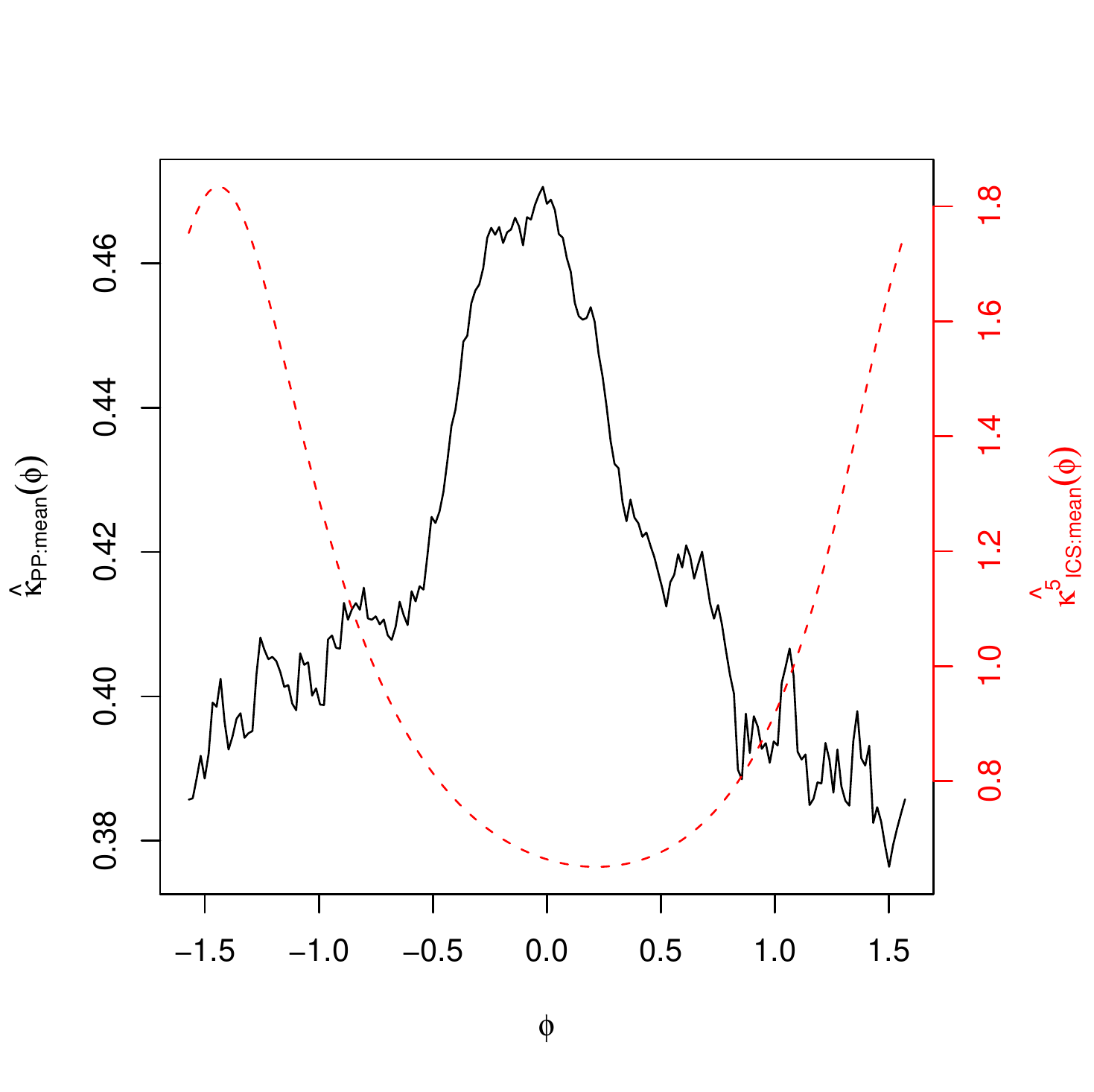}}
\caption{For $\delta=0.95$ and $q=1/2$, plots of different ICS (red dashed curve) and PP (black solid curve) criteria without (left) and with imposing a common location measure (right).}
\label{fig:fig5}
\end{figure}

From the panels in Figure \ref{fig:fig5}, we make the following remarks based on the simulated data set:
 \begin{itemize}
 \item[(1)] Panel (a) shows that ICS:var:t2 and PP:var:t2 work well
   since $\bar{\/y}$ and $\bar{\/y_{t2}}$ are approximately equal. Hence,
   imposing a common location measure has little effect, as shown in
   (b).
 \item[(2)] Panels (c), (e), (g), (i) show examples when ICS and/or PP
   go wrong because of the difference in the location measures.
 \item[(3)] Using a common location measure fixes the problem in panel
   (d) for (PP, ICS):var:mcd, panel (f) for (PP, ICS):var:mve, and
   panel(h) for (PP, ICS):t2:mcd.
\item[(4)] From panel (j), using a common location measure in
  PP:t2:mve:mean does not seem to work well. The reason might
  be due to the unstable behaviour of the mve and lshorth.
\item[(5)] The plots generally suggest that PP will be more accurate than ICS,
  since the PP plots are narrower at the clustering direction than the
  ICS plot. This property has been confirmed empirically in \citet{Fatimah} 
for certain multivariate normal mixture models and choices of scatter matrix.
\item[(6)] Similar patterns are seen with most simulated data sets
  from this model.
\end{itemize}

\subsection*{Behaviour of  ICS:var:mcd}
\begin{figure}[H]
\centering
\subfloat[]
{\includegraphics[width=6cm,height=6cm]{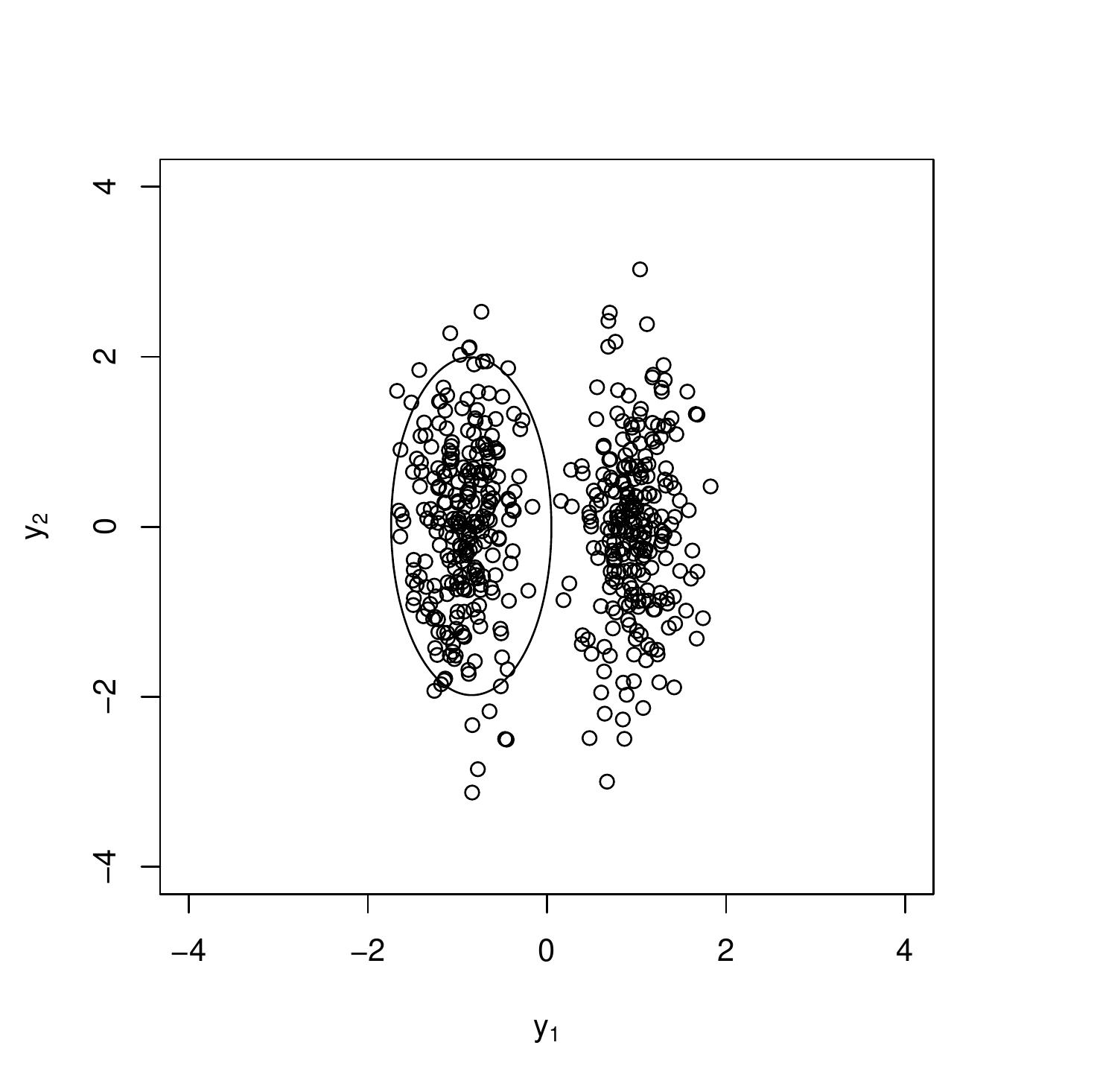}}
\subfloat[]
{\includegraphics[width=6cm,height=6cm]{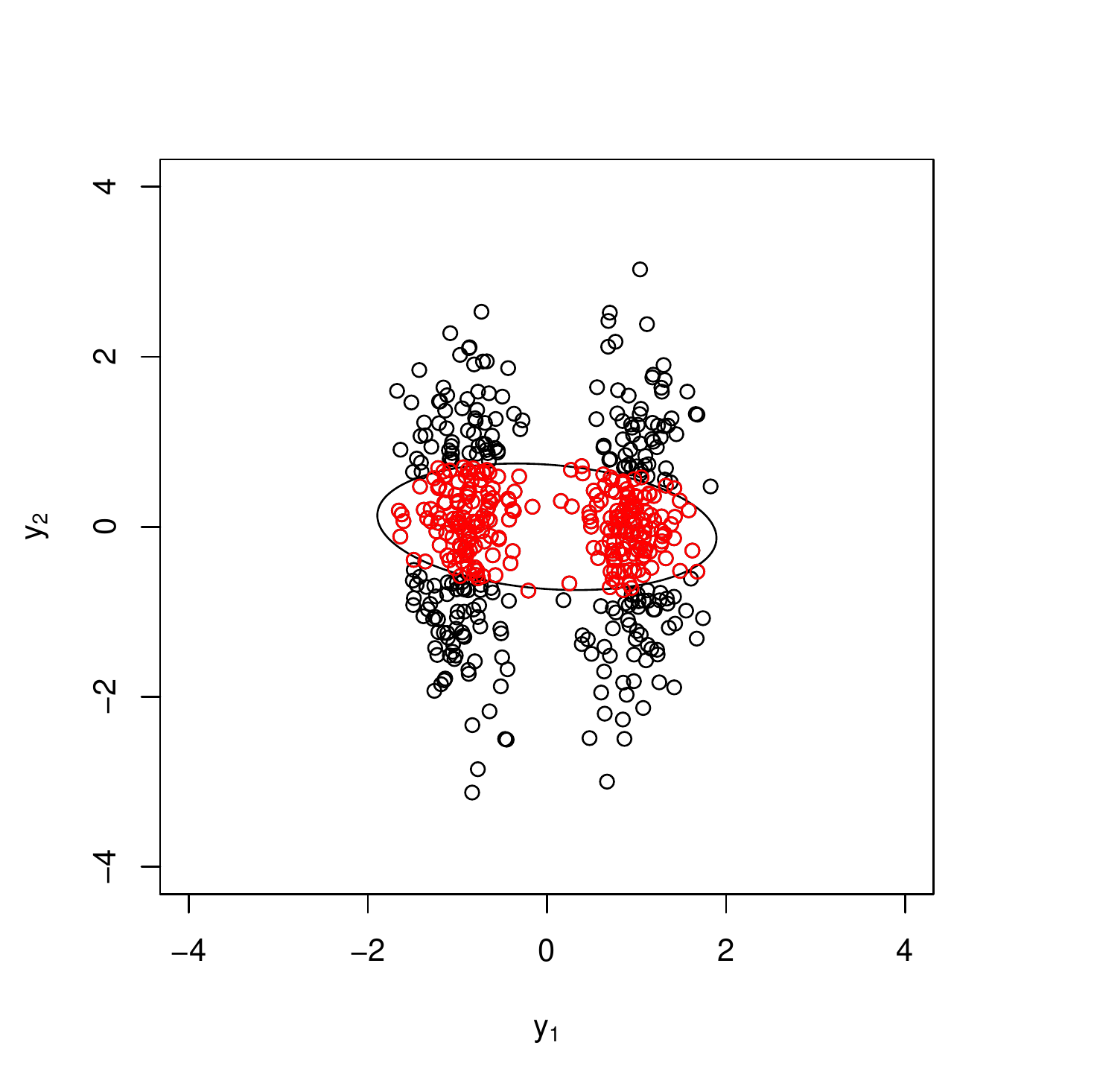}}
\caption{Plots of the ellipses of mcd scatter matrix based on (a) mcd location measure, and (b) the sample mean, superimposed on data of size $n=500$, distributed as mixtures of two normal distributions.}
\label{fig:fig6}
\end{figure}

To gain a deeper understanding of the behaviour of ICS:var:mcd in
panel \ref{fig:fig5} (c) and the effect of forcing a common location
measure on mcd in panel (d), we plot the ellipses of $S_{\text{mcd}}$
( with and without imposing a common location meaure) and superimpose
it on the data points of our example. The plots are shown in panels
\ref{fig:fig6} (a) and (b).  The behaviour in this example agrees with
the interpretation given for the population example in Section
\ref{sec:sec5}.

\subsection*{Behaviour of PP:var:mcd}
The objective function for PP:var:mcd, has a similar problem to ICS;
it is maximized rather than minimized near the
correct clustering direction.

\begin{figure}[h!]
\centering
\subfloat[ $v_{\trunc}=0.08$, $\bar{x}_{\trunc}= -0.93$]
{\includegraphics[width=0.4 \columnwidth, height=4cm ]{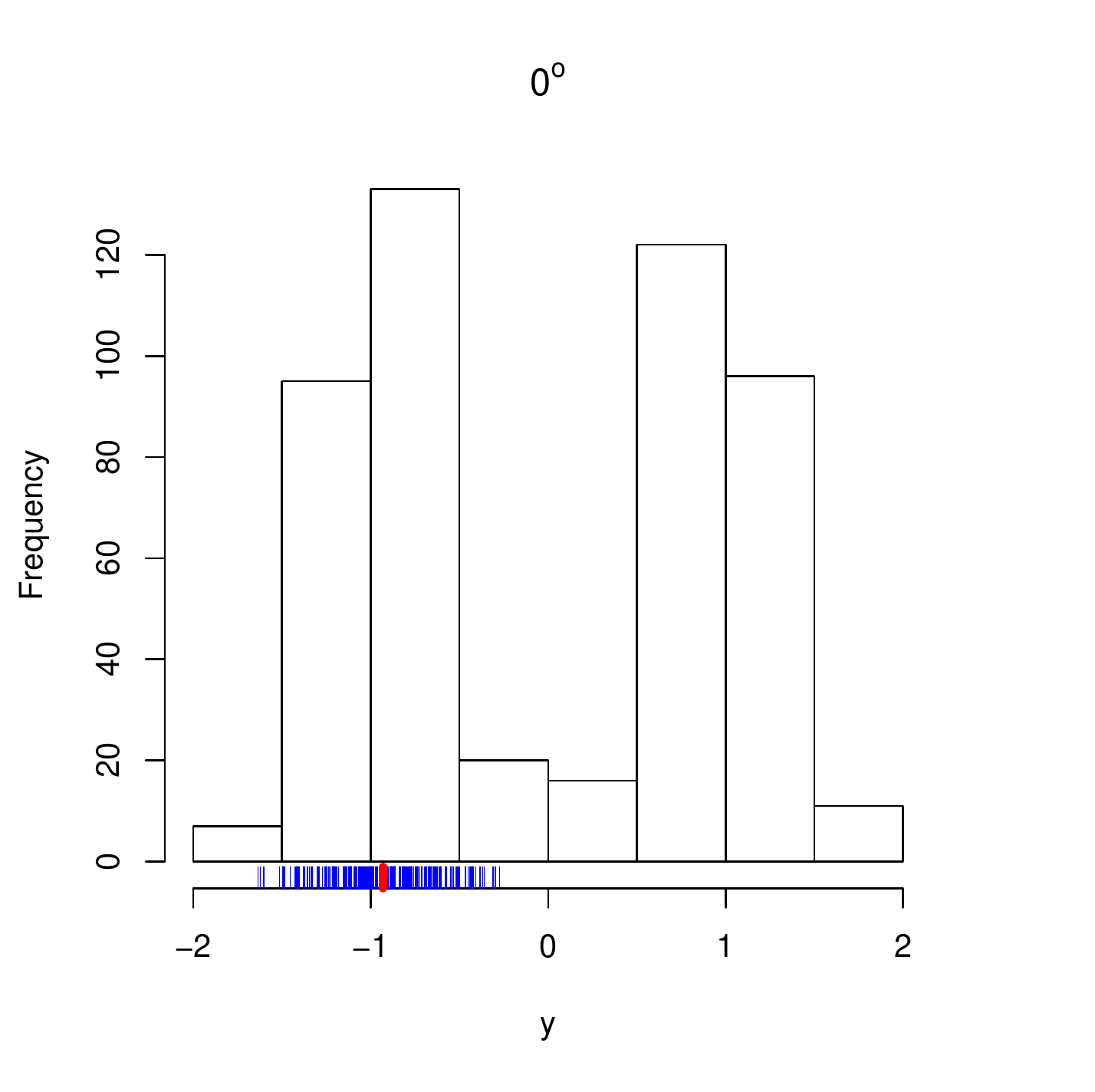}} 
\subfloat[ $v_{\trunc}=0.54$, $\bar x=0$ ]{\includegraphics[width=0.4 \columnwidth, height=4cm ]{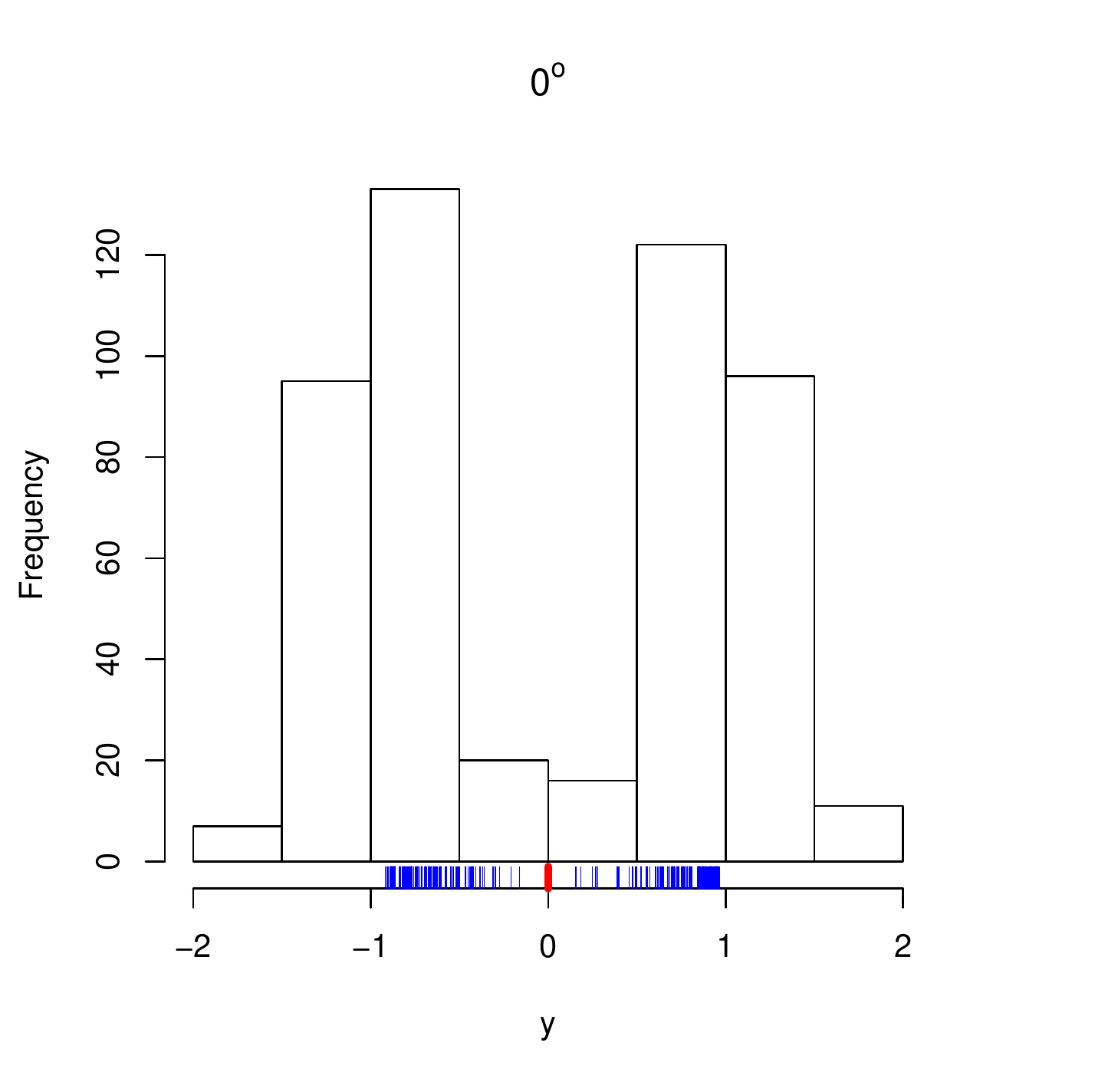}} 
\\[-3ex]
\subfloat[ $v_{\trunc}=0.14$, $\bar{x}_{\trunc}=-0.92 $]{\includegraphics[width=0.4 \columnwidth, height=4cm ]{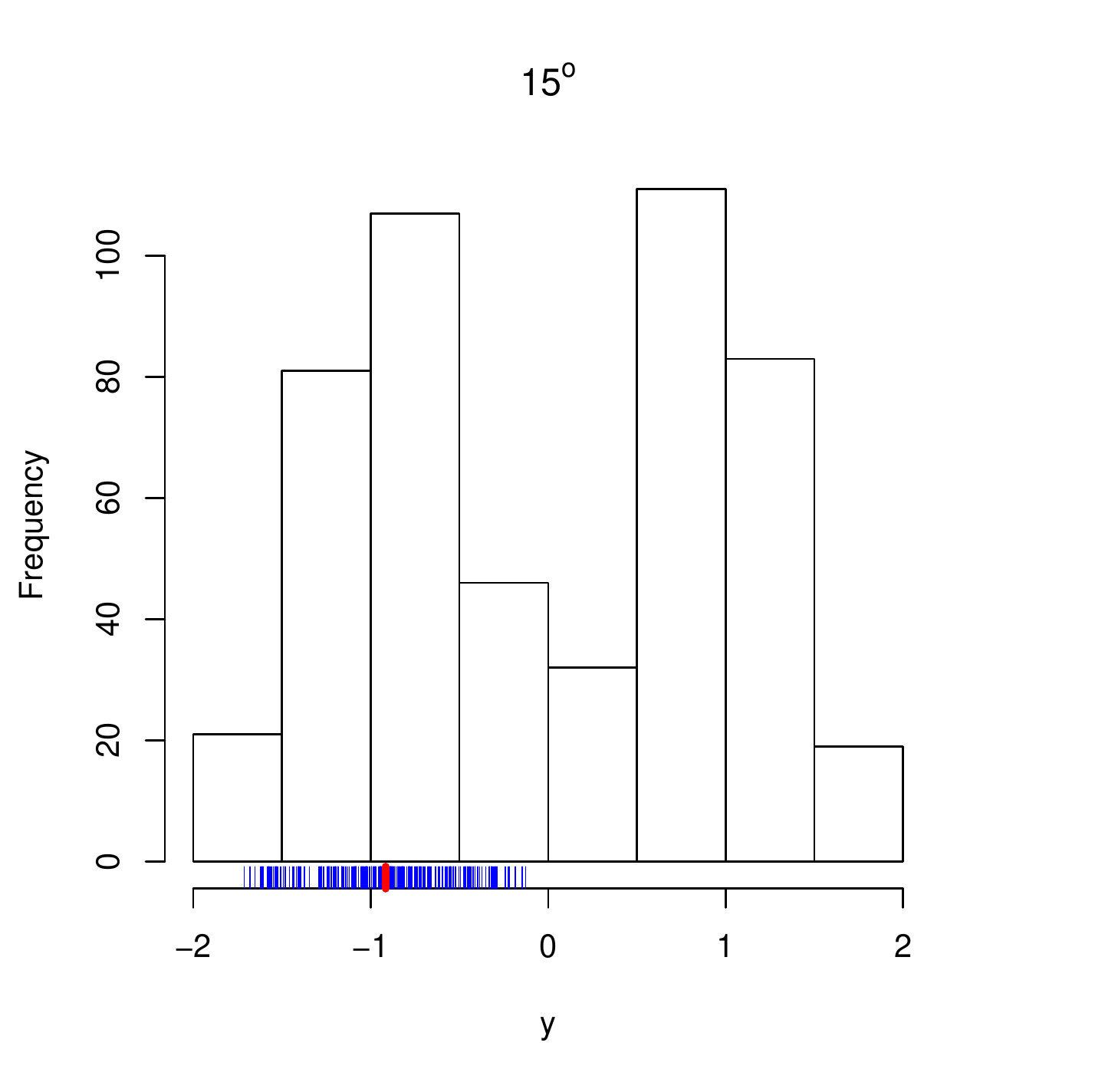}}
\subfloat[$v_{\trunc}=0.42$, $\bar x= 0$ ]{\includegraphics[width=0.4 \columnwidth, height=4cm ]{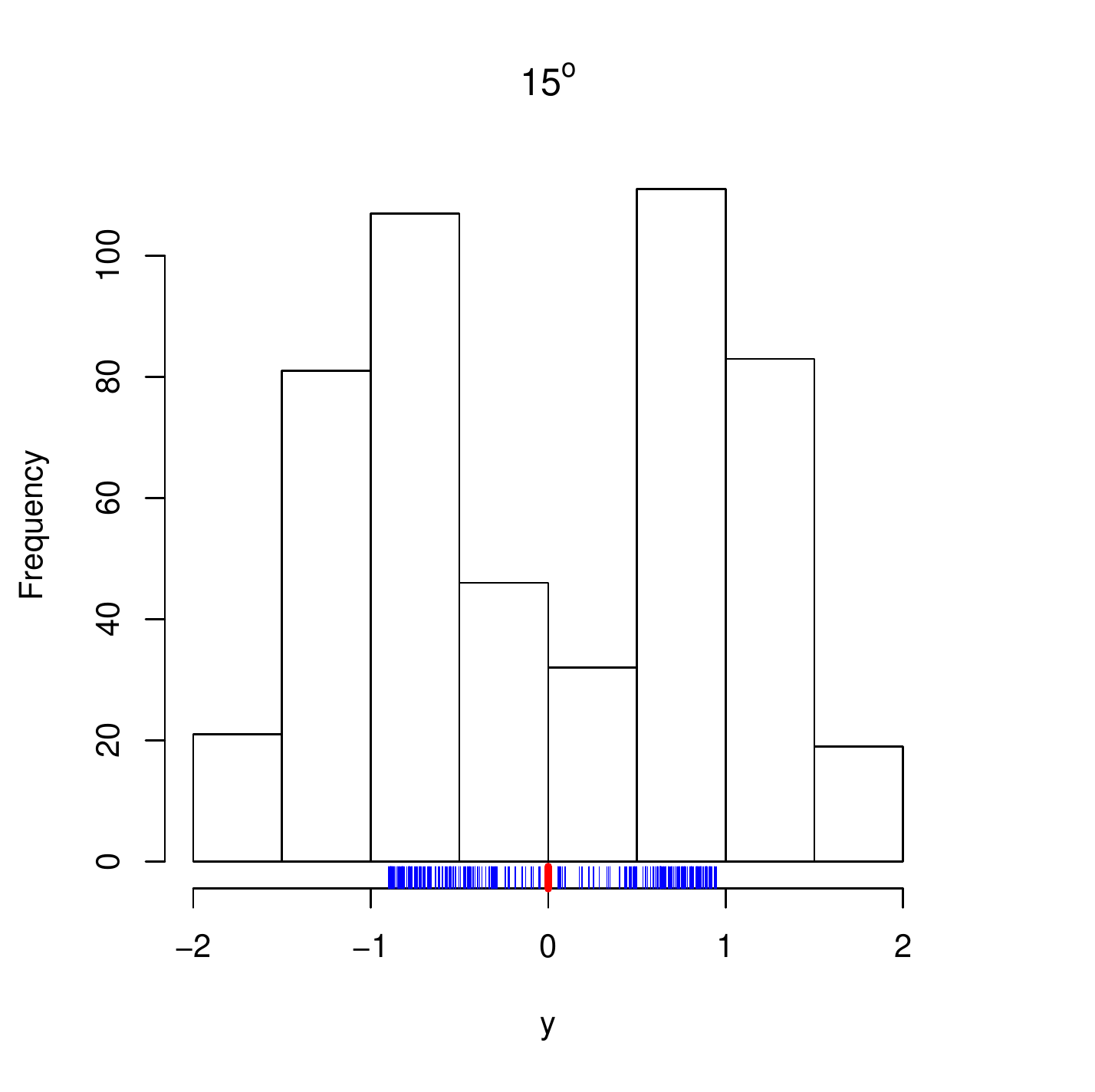}} 
\\[-3ex]
\subfloat[ $v_{\trunc}=0.21 $, $\bar{x}_{\trunc}=0.63$]{\includegraphics[width=0.4 \columnwidth, height=4cm ]{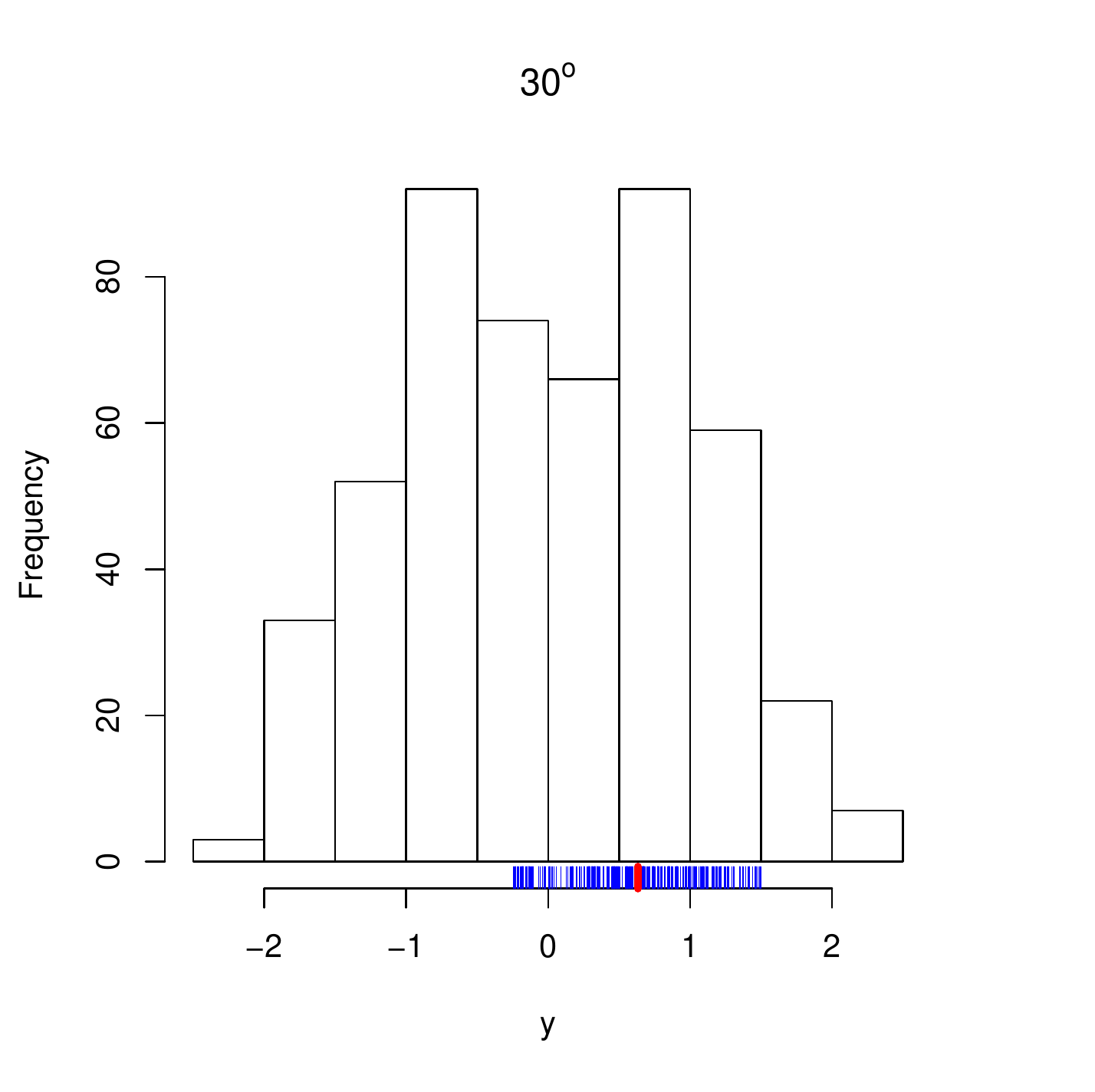}}
\subfloat[ $v_{\trunc}= 0.26$, $\bar{x}_{\trunc}= 0.02$]{\includegraphics[width=0.4 \columnwidth, height=4cm ]{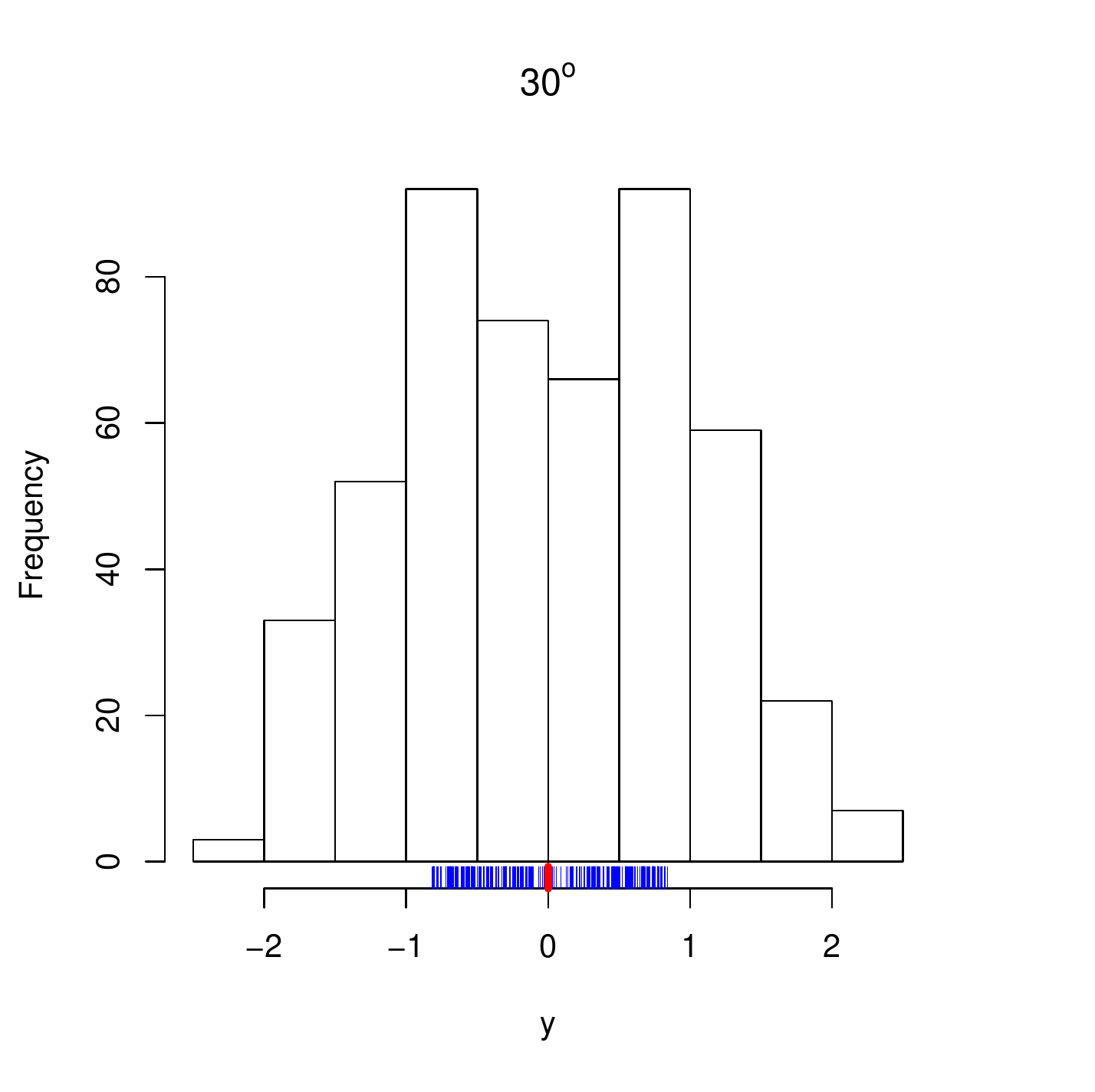}} 
\\[-3ex]
\subfloat[ $v_{\trunc}=0.12 $, $\bar{x}_{\trunc}=0.18 $]{\includegraphics[width=0.4 \columnwidth, height=4cm]{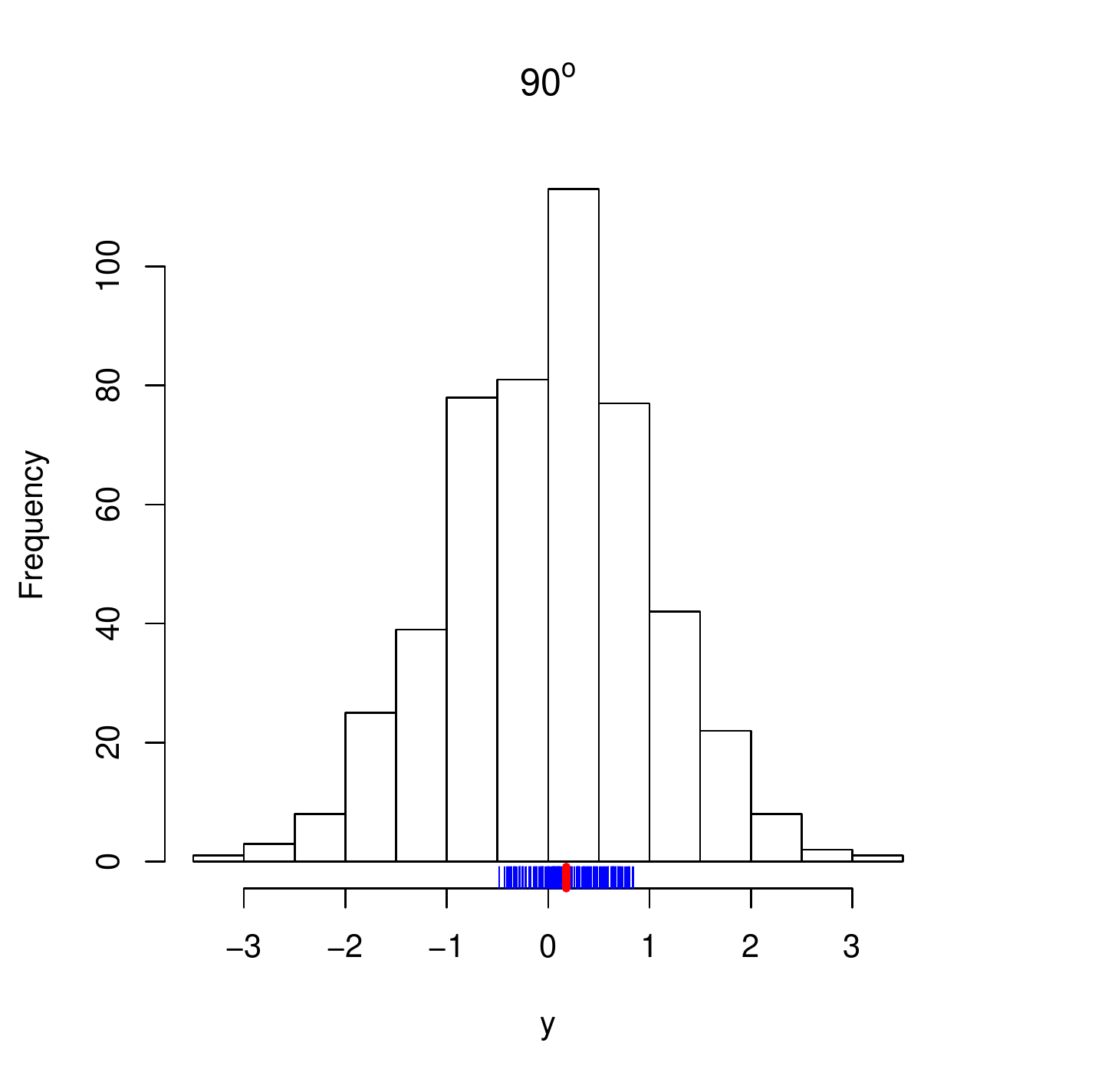}}
\subfloat[ $v_{\trunc}=0.14 $, $\bar{x}_{\trunc}= 0$]{\includegraphics[width=0.4 \columnwidth, height=4cm]{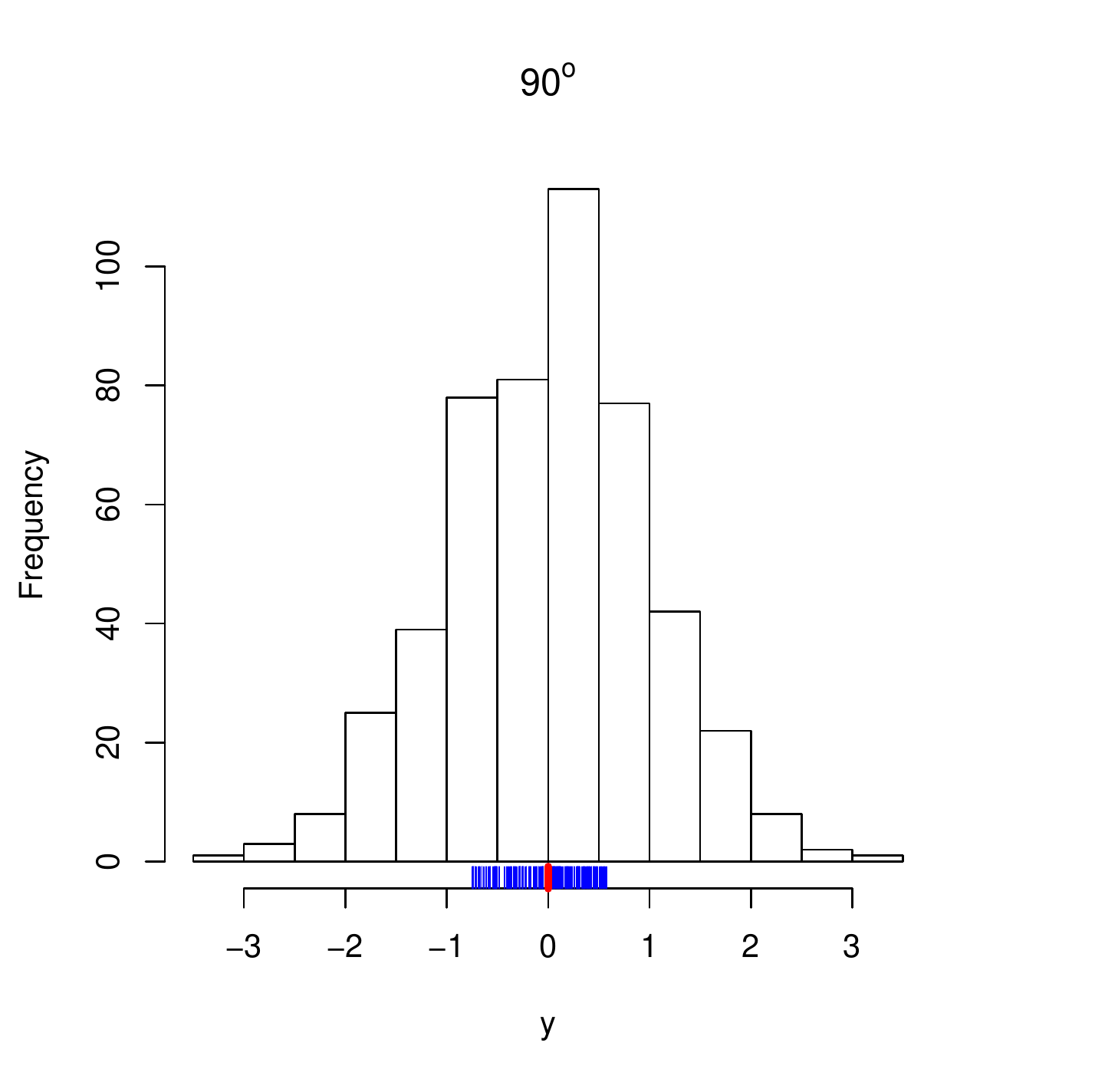}} 
\caption{Histograms of $0^\circ, 15^\circ, 30^\circ$ and $90^\circ$ projections. Left panels show
the vectors of $50\%$ of data with the smallest variance (the blue lines), and its 
location measure (the red lines), right panels show the $50\%$ of data with the smallest 
variance computed around the mean 0.}
\label{fig:fig8}
\end{figure}

To understand this behaviour in more detail, we plot in Figure \ref{fig:fig8}
one-dimensional histograms after projections by the following choices
for the angle $\phi$: $0^\circ$, $15^\circ$, $30^\circ$, and
$90^\circ$.  For each histogram, we plot the $50\%$ of the data that
has the smallest variance, and the corresponding location measure
$\bar{x}_{\trunc}$. The plots are repeated where the location measure is
constrained at the sample mean $\bar x=0$. 
\newpage

The shape of the histograms depends on of the projection directions. Note that
as $v_{\trunc}$ gets smaller, the PP criterion $\kappa_{\text{PP}}$  gets larger.

\begin{itemize}
\item[(1)] The $0^\circ$ projection produces two widely separated
  groups with one group is slightly bigger than the other.  In this
  case, $\bar{x}_{\trunc}$ is at the larger group and $v_{\trunc}$ is essentially
  the variance of this group.  Hence $v_{\trunc}$ takes its smallest value
  and $\kappa_{\text{PP}}$ is largest.

\item[(2)] The $15^\circ$ projection produces two slightly separated
  groups with within-group variance is larger than in the $0^\circ$
  projection. The value of $v_{\trunc}$ is larger than for $0^\circ$.
\item[(3)] The $30^\circ$ projection produces one group, with a
  pseudo-uniform distribution. The value of $v_{\trunc}$ is larger than for
  $15^\circ$.
\item[(4)] The $90^\circ$ projection produces one normally distributed
  group.  The value for $v_{\trunc}$ becomes small again.
\end{itemize}

Constraining the mean to be at the origin fixes the problem.  The value of
$v_{\trunc}$ steadily decreases from $0^\circ$ to $90^\circ$.

\section*{Appendix}
\renewcommand{\theequation}{A.\arabic{equation}}
\setcounter{equation}{0}
Consider the limiting balanced bivariate normal mixture model,
$$
\/y = s \/e_1 + z \/e_2,
$$
where $s = \pm 1$, each with probability 1/2, independent of $z \sim
N(0,1)$, and $\/e_1 = (1,0)^T$, $\/e_2 = (0,1)^T$.  This model is
standardized with respect to the ``total'' coordinates; i.e. $E(\/y) =
\/0$ and $\var(\/y) = I_2$. The model can also be described in terms
of a mixture of two normal distributions, concentrated on the vertical
lines $y_1=1$ and $y_1 = -1$.

In this appendix we shall show that the population version of the mve,
constrained to be centred at at the origin, is given by
$$
\Sigma_{\text{mve}} = \begin{bmatrix} 2 & 0 \\ 0 & d^2 \end{bmatrix},
$$
where $d = \Phi^{-1}(.75)$ in terms of the cumulative distribution function
of the $N(0,1)$ distribution.

First let $u_1 < u_2$ be two possible values
for $y_2$ and consider and ellipse based on a matrix $\Sigma$, with inverse
$\Sigma^{-1} = \Omega$,
\begin{equation}
\label{A:ellipse}
\/y^T \Omega \/y = 1
\end{equation}
which intersects the vertical lines at these points,
\begin{equation}
\label{A:ellipse-lines}
\begin{bmatrix} 1 & u_1 \end{bmatrix} \Omega 
\begin{bmatrix} 1 \\ u_1\end{bmatrix} = 1, \quad 
\begin{bmatrix} 1 & u_2 \end{bmatrix} \Omega 
\begin{bmatrix} 1 \\ u_2\end{bmatrix} = 1.
\end{equation}
By symmetry the ellipse also intersects the points $(-1,-u_1)^T$ and
$(-1,-u_2)^T$.  Note that $\Sigma$ will be a candidate for the mve
matrix if the interior of the ellipse covers 50\% of the probability
mass, that is,
\begin{equation}
\label{A:coverage}
\Phi(u_2) = \Phi(u_1)+1/2.
\end{equation}
If $u_1$ and $u_2$ are finite, then necessarily $u_1 < 0$ and $u_2 >0$.

The proof will proceed in two stages.  First, for fixed $u_1, \ u_2$
satisfying \eqref{A:coverage}, we choose $\Sigma$ to minimize
$\det(\Sigma)$ (or equivalently maximize $\det(\Omega)$).  Secondly,
we optimize over the choice of $u_1, \ u_2$.

Thus, start with a fixed pair of values $u_1, \ u_2$
satisfying \eqref{A:coverage}.  If $\/y = (1,u)^T$ represents a point on one of the vertical lines, then the intersection with the ellipse \eqref{A:ellipse} can be written
$$
\omega_{11} + 2 \omega_{12} u + \omega_{22} u^2 = 1,
$$
or equivalently as  the quadratic equation in u,
$$
A u^2 + B u + C = 0,
$$
where $A = \omega_{22}$, $B = 2 \omega_{12}$, $C = \omega_{11} -1$.
If this ellipse passes through $(1,u_1)^T$ and $(1,u_2)^T$, then then
$u_1, \ u_2$ are roots of the quadratic equation, so
\begin{equation}
\label{A:quad-roots}
u_1,u_2 = \frac{-B \pm \sqrt{B^2 - 4AC}}{2A}.
\end{equation}
In particular, setting $M = (u_1+u_2)/2$ to be the mean of the roots,
and $P = u_1u_2$ to be the product of the roots, we have
\begin{equation}
\label{A:roots}
M = -\frac{B}{2A} = -\frac{\omega_{12}}{\omega_{22}}, \quad
P = \frac{C}{A} = \frac{\omega_{11} - 1}{\omega_{22}}.
\end{equation}

Let us try to maximize $\det(\Omega)$ subject to the ellipse 
satisfying \eqref{A:ellipse-lines}.  Start with an arbitrary  $\omega_{22}>0$.
Then \eqref{A:roots} determines the remaining elements of $\Omega$,
$$
\omega_{12} = -M \omega_{22}, \quad \omega_{11} = 1+P \omega_{22}.
$$
Hence 
$$
\det(\Omega) = \omega_{11} \omega_{22} - \omega_{12}^2 = 
\omega_{22} - Q \omega_{22}^2,
$$
where 
\begin{equation}
\label{A:Q}
Q = M^2 - P = \frac{1}{4}(u_1 - u_2)^2> 0.
\end{equation}
Maximizing $\det(\Omega)$ with respect to the choice of $\omega_{22}$ 
leads to $\omega_{22} = 1/(2Q)$ and
$$
\det(\Omega) = 1/(4Q).
$$

The remaining task is to choose $u_1<0$ (which determines $u_2>0$ by 
\eqref{A:coverage}) to maximize $\det(\Omega)$, or equivalently, to 
minimize $Q$ in \eqref{A:Q}.  

Recall a basic result from calculus.  If $t=f(u)$ and $u = g(t)$ are
monotone functions which are inverse to one another, then $g(f(u)) =
u$. Differentiating two times yields the relation between the
derivatives,
$$
g' = 1/f', \quad g'' = -f''/\{f'\}^3.
$$
In particular, consider $f(u) = \Phi(u)$, with derivatives $f'(u) =
\phi(u)$ and $f''(u) = -u \phi(u)$, where $\phi(u)$ is the probability
density function of $N(0,1)$. Then $g(t) = \Phi^{-1}(t)$ with
derivatives $g'(t) = 1/\phi(u)$ and $g''(t) = u/\{\phi(u)\}^2$, where
$u = \Phi^{-1}(t)$.

With this notation, write $u_1 = g(t)$ for $0 < t < 1/2$. Then $u_2 =
g(t+1/2)$.  Write $\phi_1 = \phi(u_1), \ \phi_2 = \phi(u_2)$.  The
quantity $Q$ in \eqref{A:Q}, treated as a function of $t$, has
derivatives
\begin{align*}
Q' &= \frac{1}{2} \left\{ u_1 u_1' - u_1 u_2' - u_1'u_2 + u_2 u_2'\right\}\\
&= \frac{1}{2} \left\{u_1(1/\phi_1 -1/\phi_2) + 
                      u_2(1/\phi_2 -1/\phi_1) \right\}\\
Q'' &= \frac{1}{2} \left\{ u_1 u_1''+ (u_1')^2 -u_1 u_2'' - 2 u_1' u_2' - 
               u_1''u_2 + u_2 u_2''+ (u_2')^2 \right\}\\
    &= \frac{1}{2} \left\{ u_1^2/\phi_1^2 + 1/\phi_1^2 - u_1u_2/\phi_2^2 -
2/(\phi_1 \phi_2) - u_1u_2/\phi_1^2 + u_2^2/\phi_2^2 + 1/\phi_2^2 \right\}\\
&= \frac{1}{2} \left\{ (1/\phi_1 - 1/\phi_2)^2 + 
u_1^2/\phi_1^2 + - u_1u_2/(\phi_1^2 + \phi_2^2) + u_2^2/\phi_2^2 \right\}.
\end{align*}
If $u_1=-d$, then $u_2=d$ and $\phi_1 = \phi_2$ so that the first
derivative vanishes.  For all $(0 < t< 1/2)$, the second derivative is
positive, so the function is convex.  Hence $Q$ is minimized for
$u_1=-d, \ u_2=d$.  Then $M=0, Q=-P = d^2$ and the optimal $\Sigma$
becomes
$$
\Sigma = \Omega^{-1} = \begin{bmatrix} 2 & 0 \\ 0 & 2 d^2 \end{bmatrix},
$$
as required.
\clearpage
\bibliographystyle{apalike}
\bibliography{paper_ref3}
\end{document}